\documentclass[journal]{IEEEtran}

\usepackage{amssymb,epic,eepic}
\usepackage[dvips]{hyperref}
\usepackage[dvips]{color}
\usepackage{graphicx}
\usepackage{pst-plot, pstricks}
\usepackage{fancybox,amssymb,color}

\begin{document}
\title{Stationary Algorithmic Probability}

\newcommand{\cn}{{\left(\mathbb{C}^2\right)^{\otimes n}}}
\newcommand{\cnq}{{\left(\mathbb{C}^2\right)^{\otimes n}_{\mathbb{Q}}}}
\newcommand{\hr}{{\cal H}}
\newcommand{\bo}{{\cal B}}
\newcommand{\C}{{\mathbb C}}
\newcommand{\R}{{\mathbb R}}
\newcommand{\N}{{\mathbb N}}
\newcommand{\idn}{\mathbf{1}}
\newcommand{\Z}{{\mathbb Z}}
\newcommand{\x}{{\mathbf x}}
\newcommand{\eps}{{\varepsilon}}
\newcommand{\vphi}{{\varphi}}
\newcommand{\om}{\omega}
\newcommand{\kk}{{\mathbf k}}
\newcommand{\n}{{\mathbf n}}
\newcommand{\y}{{\mathbf y}}
\newtheorem{theorem}{Theorem}[section]
\newtheorem{lemma}[theorem]{Lemma}
\newtheorem{corollary}[theorem]{Corollary}
\newtheorem{example}[theorem]{Example}
\newtheorem{definition}[theorem]{Definition}
\newtheorem{proposition}[theorem]{Proposition}
\newtheorem{conjecture}[theorem]{Conjecture}
\newcommand{\nix}{{\rule{0pt}{2pt}}}
\newcommand{\qedd}{{\nix\nolinebreak\hfill\hfill\nolinebreak$\Box$}}
\newcommand{\qed}{{\qedd\par\medskip\noindent}}
\newcommand{\lineclear}{{\rule{0pt}{0pt}\nopagebreak\par\nopagebreak\noindent}}
\newcommand{\s}{{\{0,1\}^*}}
\newcommand{\sbar}{{\overline\s}}
\newcommand{\bars}{{\overline\s}}
\newcommand{\emu}[1]{\stackrel{#1}{\longrightarrow}}
\newcommand{\nemu}[1]{\stackrel{#1}{\not\longrightarrow}}

\author{Markus M\"uller%
\thanks{M. M\"uller is with the Institute of Mathematics, Technical University of Berlin (e-mail:
mueller@math.tu-berlin.de).}}%
\markboth{Stationary Algorithmic Probability (February 4, 2009)}{}

\maketitle

\begin{abstract}
Kolmogorov complexity and algorithmic probability are defined only up to an additive resp. multiplicative constant,
since their actual values depend on the choice of the universal reference computer. In this paper, we analyze a natural approach to
eliminate this machine-dependence.

Our method is to assign algorithmic probabilities to the different computers themselves, based on the
idea that ``unnatural'' computers should be hard to emulate. Therefore, we study the Markov process of
universal computers randomly emulating each other. The corresponding stationary distribution, if it existed, would give a natural
and machine-independent
probability measure on the computers, and also on the binary strings.

Unfortunately, we show that no stationary distribution exists on the set of all computers; thus, this method
cannot eliminate machine-dependence. Moreover, we show that the reason for failure has a clear and
interesting physical interpretation, suggesting that every other conceivable attempt to get rid of those additive
constants must fail in principle, too.

However, we show that restricting to some subclass of computers might help to get rid of
some amount of machine-dependence in some situations, and the resulting stationary computer and string probabilities have
beautiful properties.
\end{abstract}


\IEEEpeerreviewmaketitle

\section{Introduction and Main Results}
\label{SecIntro}
\PARstart{S}{ince} algorithmic probability has first been studied in the 1960s
by Solomonoff, Levin, Chaitin
and others (cf. \cite{Solomonoff}, \cite{Levin}, \cite{Chaitin}), it has revealed
a variety of interesting properties, including applications in
computer science, inductive inference and statistical mechanics (cf. ~\cite{LiVitanyi}, \cite{Discovery},
\cite{Hutter}).
The algorithmic probability of a binary string $s$ is defined as the probability that
a universal prefix computer $U$ outputs $s$ on random input, i.e.
\begin{equation}
   P_U(s):=\sum_{x\in\s:U(x)=s} 2^{-|x|},
   \label{EqDefAP}
\end{equation}
where $|x|$ denotes the length of a binary string $x\in\s$. It follows from the Kraft inequality that
\[
   \sum_{s\in\s} P_U(s)=:\Omega_U< 1,
\]
where $\Omega_U$ is Chaitin's famous halting probability. So algorithmic probability
is a subnormalized probability distribution or semimeasure on the binary strings. It is closely related to prefix Kolmogorov
complexity $K_U(s)$ which is defined~\cite{LiVitanyi} as the length of the shortest computer program that outputs $s$:
\[
   K_U(s):=\min\{|x|\,\,|\,\, U(x)=s\}.
\]
The relation between the two can be written as
\begin{equation}
   K_U(s)=-\log P_U(s)+\mathcal{O}(1),
   \label{EqRelKP}
\end{equation}
where the $\mathcal{O}(1)$-term denotes equality up to an additive constant. Both Kolmogorov complexity and algorithmic
probability depend on the choice of the universal reference computer $U$. However, they do not depend on $U$ ``too much'':
If $U$ and $V$ are both universal prefix computers, then it follows from the fact that one can emulate the other that
\[
   K_U(s)=K_V(s)+\mathcal{O}(1),
\]
i.e. the complexities $K_U$ and $K_V$ differ from each other only up to an additive constant. Then Equation~(\ref{EqRelKP})
shows that the corresponding algorithmic probabilities differ only up to a multiplicative constant.

This kind of ``weak'' machine independence is good enough for many applications: if the strings are long enough,
then a fixed additive constant does not matter too much. However, there are many occasions where it would be desirable
to get rid of those additive constants, and to eliminate the arbitrariness which comes from the choice of the
universal reference computer. Examples are Artificial Intelligence~\cite{Hutter} and physics~\cite{Schack}, where one
often deals with finite and short binary strings.

We start with a simple example, to show that the machine-dependence of algorithmic probability can be drastic,
and also to illustrate the main idea of our approach. Suppose that $U_{nice}$ is a ``natural'' universal prefix computer,
say, one which is given by a Turing machine model that we might judge as ``simple''. Now choose
an arbitrary strings $s$ consisting of a million random bits; say, $s$ is attained by a million tosses of a fair coin.
With high probability, there is no short program for $U_{nice}$ which computes $s$ (otherwise toss the coin again
and use a different string $s$). We thus expect that
\[
   P_{U_{\rm nice}}(s)\approx 2^{-1.000.000}.
\]
Now we define another prefix computer $U_{\rm bad}$ as
\[
   U_{\rm bad}(x):=\left\{
      \begin{array}{cl}
         s & \mbox{if }x=0,\\
         {\rm undefined} & \mbox{if }x=\lambda\mbox{ or }x=0y,\\
         U_{\rm nice}(y) & \mbox{if }x=1y.
      \end{array}
   \right.
\]
The computer $U_{\rm bad}$ is universal, since it emulates the universal computer $U_{\rm nice}$
if we just prepend a ``1'' to the input. Since $U_{\rm bad}(0)=s$, we have
\[
   P_{U_{\rm bad}}(s)> \frac 1 2.
\]
Hence the algorithmic probability $P_U(s)$ depends drastically on the choice of the universal computer $U$.
Clearly, the computer $U_{\rm bad}$ seems quite unnatural,
but in algorithmic information theory, all the universal computers are created equal --- there is
no obvious way to distinguish between them and to say which one of them is a ``better choice'' than the other.

So what {\em is} ``the'' algorithmic probability of the single string $s$? It seems clear
that $2^{-1.000.000}$ is a better answer than $\frac 1 2$, but the question is how we can
make mathematical sense of this statement. 
How can we give a sound formal meaning to the statement that $U_{\rm nice}$ is more ``natural''
than $U_{\rm bad}$? A possible answer is that in the process of randomly constructing a computer from scratch,
one is very unlikely to end up with $U_{\rm bad}$,
while there is some larger probability to encounter $U_{\rm nice}$.

This suggests that we might hope to find some natural probability distribution
$\mu$ on the universal computers, in such a way that $\mu(U_{\rm bad})\ll\mu(U_{\rm nice})$.
Then we could define the ``machine-independent'' algorithmic probability $P(s)$ of some string $s$
as the weighted average of all algorithmic probabilities $P_U(s)$,
\begin{equation}
   P(s):=\sum_{U\mbox{ universal}}\mu(U) P_U(s).
   \label{EqProbIdea}
\end{equation}
Guided by Equation~(\ref{EqRelKP}), we could then define ``machine-independent Kolmogorov complexity''
via $K(s):=-\log P(s)$.

But how can we find such a probability distribution $\mu$ on the computers? The key idea here is to compare
the capabilities of the two computers to emulate each other. Namely,
by comparing $U_{\rm nice}$ and $U_{\rm bad}$, one observes that
\begin{itemize}
\item it is very ``easy'' for the computer $U_{\rm bad}$ to emulate the computer $U_{\rm nice}$:
just prepend a ``1'' to the input. On the other hand,
\item it is very ``difficult'' for the computer $U_{\rm nice}$ to emulate $U_{\rm bad}$:
to do the simulation, we have to supply $U_{\rm nice}$ with the long string $s$ as
additional data.
\end{itemize}
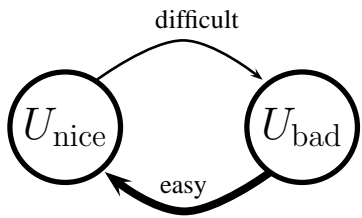
\begin{figure}[!hbt]
\begin{center}
\psset{unit=0.35cm}
\begin{pspicture}(0,2.5)(12.3,9)
  \rput(3,5.2){\LARGE $U_{\rm nice}$}
  \pscircle[linewidth=0.2](3,5.2){2.2}
  \rput(12,5.2){\LARGE $U_{\rm bad}$}
  \pscircle[linewidth=0.2](12,5.2){2.2}
  {\psset{linewidth=0.1}\pscurve{->}(4.2,7)(7.5,8.5)(10.5,7)}
  \rput(0,0){}
  {\psset{linewidth=0.3}\pscurve{<-}(4.5,3.5)(7.5,2)(10.8,3.5)}
  \rput(8,9.3){difficult}
  \rput(7.5,2.8){easy}
\end{pspicture}
\end{center}
\caption{Computers that emulate each other.}
\label{FigEasyDiff}
\end{figure}
The idea is that this observation holds true more generally: {\em ``Unnatural'' computers are harder to emulate.}
There are two obvious approaches to construct some computer probability $\mu$ from this observation --- interestingly,
both turn out to be equivalent:
\begin{itemize}
\item The situation in Figure~\ref{FigEasyDiff} looks like the graph of some
Markov process. If one starts with either one of the two computers depicted there
and interprets the line widths as transition probabilities, then in the long
run of more and more moves, one tends to have larger probability to end up at $U_{\rm nice}$
than at $U_{\rm bad}$. So let's apply this idea more generally and define a Markov process of all the
universal computers, randomly emulating each other. If the process has a stationary distribution (e.g. if it is positive recurrent),
this is a good candidate for computer probability.
\item Similarly as in Equation~(\ref{EqDefAP}), there should be a simple way to define probabilities $P_U(V)$
for computers $U$ and $V$, that is, the probability that $U$ emulates $V$ on random input.
Then, whatever the
desired computer probability $\mu$ looks like, to make any sense, it should satisfy
\[
   \mu(U)=\sum_{V\mbox{ universal}} \mu(V) P_V(U).
\]
But if we enumerate all universal computers as $\{U_1,U_2,U_3,\ldots\}$, this equation can be written as
\[
   \left(\begin{array}{c}\mu(U_1)\\ \mu(U_2) \\ \mu(U_3) \\ \ldots \end{array}\right)=
   \left(
      \begin{array}{cc}
         P_{U_1}(U_1) & P_{U_2}(U_1)  \ldots\\
         P_{U_1}(U_2) & P_{U_2}(U_2)  \ldots \\
         P_{U_1}(U_3) & P_{U_2}(U_3)  \ldots \\
         \ldots & \ldots \ldots
      \end{array}
   \right)\cdot
   \left(\begin{array}{c}\mu(U_1)\\ \mu(U_2) \\ \mu(U_3) \\ \ldots \end{array}\right)
\]
Thus, we should look for the unknown stationary probability eigenvector $\underline{\mu}$ of
the ``emulation matrix'' $\left( P_{U_i}(U_j)\right)_{i,j}$.
\end{itemize}
Clearly, both ideas are equivalent if the probabilities $P_U(V)$ are the transition probabilities
of the aforementioned Markov process.

Now we give a synopsis of the paper and explain our main results:\label{Synopsislabel}
\begin{itemize}
\item Section~\ref{SecPreliminaries} contains some notational preliminaries, and defines
the {\em output frequency} of a string as the frequency that this string is output by a computer.
For prefix computers, this notion equals algorithmic probability (Example~\ref{ExPrefix}).
\item In Section~\ref{SecStationaryComputerProbability}, we define the emulation Markov process that we have
motivated above, and analyze if it has a stationary distribution or not.
Here is the construction for the most important case (the case of the full set of computers)
in a nutshell: we say that a computer $C$ {\em emulates} computer $D$ via the string $x$,
and write $C\emu{x}D$ and $D=\left(C\emu{x}\right)$ if $C(xy)=D(y)$ for all strings $y$. A computer is {\em universal} if it emulates every other computer.
Given a universal computer, at least one of the two computers $C\emu{0}$ and $C\emu{1}$ must be universal, too.

Thus, we can consider the universal computers as the vertices of a graph, with directed edges going from $U$ to $V$ if
$U\emu{0}V$ or $U\emu{1}V$. Every vertex (universal computer) has either one or two outgoing edges (corresponding to
the two bits). The random walk on this connected graph defines a Markov process: we start at some computer, follow the outgoing edges,
and if there are two edges, we follow each of them with probability $\frac 1 2$. This
is schematically depicted in Figure~\ref{FigEmulationMarkovProcess}.

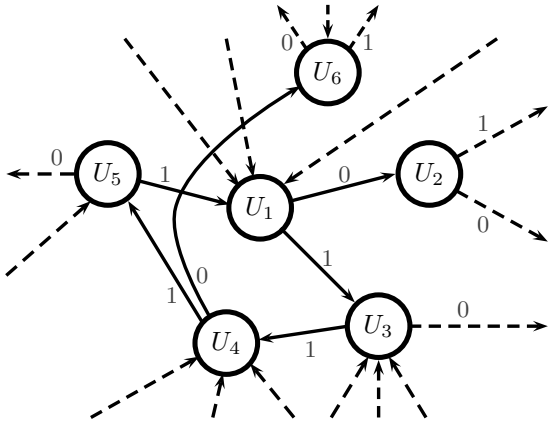
\begin{figure}[!hbt]
\begin{center}
\psset{unit=0.45cm}
\begin{pspicture}(-1,-1.5)(11,11)

   \pscircle[linewidth=0.15](5,5){1}
   \rput(5,5){$U_1$}
   \pscircle[linewidth=0.15](10,6){1}
   \rput(10,6){$U_2$}
   \pscircle[linewidth=0.15](7,9){1}
   \rput(7,9){$U_6$}
   \pscircle[linewidth=0.15](4,1){1}
   \rput(4,1){$U_4$}
   \pscircle[linewidth=0.15](8.5,1.5){1}
   \rput(8.5,1.5){$U_3$}
   \pscircle[linewidth=0.15](0.5,6){1}
   \rput(0.5,6){$U_5$}
   \psline[linewidth=0.1,linestyle=dashed]{<-}(5.7,5.7)(12,10)
   \psline[linewidth=0.1,linestyle=dashed]{<-}(4.8,5.9)(4,10)
   \psline[linewidth=0.1,linestyle=dashed]{<-}(4.4,5.6)(1,10)
   \psline[linewidth=0.1]{->}(5.9,5.2)(9,6)
   \rput(7.5,6){\small$\color{darkgray}0$}
   \psline[linewidth=0.1]{->}(5.6,4.4)(7.8,2.2)
   \rput(7,3.5){\small$\color{darkgray}1$}
   
   \psline[linewidth=0.1,linestyle=dashed]{->}(10.8,6.5)(13.5,8)
   \rput(11.6,7.5){\small$\color{darkgray}1$}
   \psline[linewidth=0.1,linestyle=dashed]{->}(10.8,5.5)(13.5,4)
   \rput(11.6,4.5){\small$\color{darkgray}0$}
   
   \psline[linewidth=0.1,linestyle=dashed]{->}(9.4,1.5)(13.5,1.5)
   \rput(11,2){\small$\color{darkgray}0$}
   \psline[linewidth=0.1]{->}(7.6,1.5)(4.95,1.1)
   \rput(6.5,0.8){\small$\color{darkgray}1$}
   \psline[linewidth=0.1,linestyle=dashed]{<-}(8.2,0.6)(7.1,-1.2)
   \psline[linewidth=0.1,linestyle=dashed]{<-}(8.5,0.5)(8.5,-1.2)
   \psline[linewidth=0.1,linestyle=dashed]{<-}(8.8,0.6)(9.9,-1.2)
   
   \pscurve[linewidth=0.1]{->}(3.5,1.8)(2.5,5)(6.2,8.6)
   \rput(3.3,3){\small$\color{darkgray}0$}
   \psline[linewidth=0.1]{->}(3.3,1.6)(1.1,5.2)
   \rput(2.4,2.4){\small$\color{darkgray}1$}
   \psline[linewidth=0.1,linestyle=dashed]{<-}(4.7,0.4)(6,-1.2)
   \psline[linewidth=0.1,linestyle=dashed]{<-}(3.9,0.1)(3.6,-1.2)
   \psline[linewidth=0.1,linestyle=dashed]{<-}(3.2,0.6)(0,-1.2)

   \psline[linewidth=0.1]{->}(1.4,5.8)(4.1,5.1)
   \rput(2.2,6){\small$\color{darkgray}1$}
   \psline[linewidth=0.1,linestyle=dashed]{->}(-0.5,6)(-2.5,6)
   \rput(-1,6.5){\small$\color{darkgray}0$}
   \psline[linewidth=0.1,linestyle=dashed]{<-}(0,5.2)(-2.5,3)

   \psline[linewidth=0.1,linestyle=dashed]{->}(6.4,9.65)(5.5,11)
   \rput(5.8,9.9){\small$\color{darkgray}0$}
   \psline[linewidth=0.1,linestyle=dashed]{<-}(7,10)(7,11)
   \psline[linewidth=0.1,linestyle=dashed]{->}(7.6,9.65)(8.5,11)
   \rput(8.2,9.9){\small$\color{darkgray}1$}
\end{pspicture}
\end{center}
\caption{Schematic diagram of the emulation Markov process. Note that the zeroes and ones represent input bits,
{\em not} transition probabilities, for example $U_1(1y)=U_3(y)$ for every string $y$. In our notation, we have for example
$U_1\emu{110}U_6$.}
\label{FigEmulationMarkovProcess}
\end{figure}

If this process had a stationary distribution, this would be a good candidate for a natural algorithmic probability
measure on the universal computers. Unfortunately, no stationary distribution exists: this Markov process is transient.

We prove this in Theorem~\ref{TheMarkoffChaney}. The idea is to construct a sequence of universal computers $M_1, M_2, M_3,\ldots$
such that $M_i$ emulates $M_{i+1}$ with high probability --- in fact, with probability turning to $1$ fast as $i$ gets large.
The corresponding part of the emulation Markov process is depicted in Figure~\ref{FigVirus}.
The outgoing edges in the upwards direction lead back to a fixed universal reference computer, which
ensures that every computer $M_i$ is universal.
\begin{figure}[!hbt]
\begin{center}
\psset{unit=0.45cm}
\begin{pspicture}(0,4)(15,8)
   \pscircle[linewidth=0.15](0,5){1}
   \rput(0,5){$M_1$}
   \pscircle[linewidth=0.15](4,5){1}
   \rput(4,5){$M_2$}
   \pscircle[linewidth=0.15](8,5){1}
   \rput(8,5){$M_3$}
   \pscircle[linewidth=0.15](12,5){1}
   \rput(12,5){$M_4$}
   \psline[linewidth=0.15]{->}(0.9,5)(3.1,5)
   \psline[linewidth=0.15]{->}(4.9,5)(7.1,5)
   \psline[linewidth=0.15]{->}(8.9,5)(11.1,5)
   \psline[linewidth=0.15]{->}(12.9,5)(15.1,5)
   \rput(1.8,5.7){$\frac 1 2$}
   \rput(5.8,5.7){$\frac 3 4$}
   \rput(9.8,5.7){$\frac 7 8$}
   \rput(13.8,5.7){$\frac {15}{16}$}
   \rput(16,4.5){$\ldots$}
   \psline[linewidth=0.15,linestyle=dashed]{->}(0.4,5.9)(3,8)
   \psline[linewidth=0.15,linestyle=dashed]{->}(4.4,5.9)(7,8)
   \psline[linewidth=0.15,linestyle=dashed]{->}(8.4,5.9)(11,8)
   \psline[linewidth=0.15,linestyle=dashed]{->}(12.4,5.9)(15,8)
   \rput(1.4,7.6){$\frac 1 2$}
   \rput(5.4,7.6){$\frac 1 4$}
   \rput(9.4,7.6){$\frac 1 8$}
   \rput(13.4,7.6){$\frac 1 {16}$}
\end{pspicture}
\end{center}
\caption{This construction in Theorem~\ref{TheMarkoffChaney} proves that the emulation Markov chain is transient, and no stationary
distribution exists (the numbers are transition probabilities).
It is comparable to a ``computer virus'' in the sense that each $M_i$ emulates ``many'' slightly modified copies of itself.}
\label{FigVirus}
\end{figure}
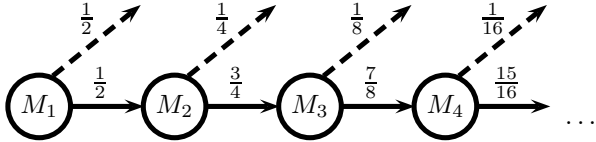

As our Markov process has only transition probabilities $\frac 1 2$ and $1$, the edges going from $M_i$ to $M_{i+1}$ in fact
consist of several transitions (edges). As those transition probabilities are constructed to tend to $1$ very fast,
the probability to stay on this $M_i$-path forever (and not return to any other computer) is positive, which forces
the process to be transient.

Yet, it is still possible to construct analogous Markov processes for restricted sets of computers $\Phi$.
Some of those sets yield processes which have stationary distributions; a non-trivial example is given in Example~\ref{ExPosRecurrent}.
\item For those computer sets $\Phi$ with positive recurrent emulation process, the corresponding
computer probability has nice properties that we study in Section~\ref{SecSymmetries}. The computer
probability induces in a natural way a probability distribution on the strings $s\in\s$ (Definition~\ref{DefStringProbability})
as the probability that the random walk described above encounters some output which equals $s$. This probability is computer-independent
and can be written in several equivalent ways (Theorem~\ref{TheStatAlgSP}).
\item A symmetry property of computer probability yields another simple and interesting proof why for the set of all computers --- and
for many other natural computer sets --- the corresponding Markov process cannot be positive recurrent (Theorem~\ref{ThePosRecSym}).
In short, if $\sigma$ is a computable permutation, then a computer $C$ and the output permuted computer
$\sigma\circ C$ must have the same probability as long as both are in the computer set $\Phi$ (Theorem~\ref{TheOutputSym}).
If there are infinitely many of them, they all must have probability zero which contradicts positive recurrence.
\item For the same reason, there cannot be one particular ``natural'' choice of a computer set $\Phi$
with positive recurrent Markov process, because $\sigma\circ\Phi$ is always another good (positive recurrent)
candidate, too (Theorem~\ref{TheNonUniqueness}).
\item This has a nice physical interpretation which we explain in Section~\ref{SecConclusions}: algorithmic probability
and Kolmogorov complexity always contain at least the ambiguity which is given by permuting the output strings.
This permutation can be interpreted as ``renaming'' the objects that the strings are describing.
\end{itemize}
We argue that this kind of ambiguity will be present in any attempt to eliminate machine-dependence from algorithmic
probability or complexity, even if it is different from the approach in this paper.
This conclusion can be seen as the main result of this work.

Finally, we show in the appendix that the string probability that we have constructed
equals, under certain conditions, the weighted average of output frequency --- this is
a particularly unexpected and beautiful result (Theorem~\ref{TheEqDef}) which needs some technical steps to be proved.
The main tool is the study of input transformations, i.e., to permute the strings before the computation.
The appendix is the technically most difficult part of this paper and can be skipped on first reading.

\section{Preliminaries and Output Frequency}
\label{SecPreliminaries}
We start by fixing some notation. In this paper, we only consider {\em finite, binary
strings}, which we denote by
\[
   \s:=\bigcup_{n=0}^\infty \{0,1\}^n=\{\lambda,0,1,00,01,\ldots\}.
\]
The symbol $\lambda$ denotes the {\em empty string}, and we write the {\em length}
of a string $s\in\s$ as $|s|$, while the cardinality of a set $S$ is denoted $\# S$. To avoid confusion with the
composition of mappings, we denote the {\em concatenation} of strings with the symbol $\otimes$, e.g.
\[
   101\otimes 001=101001.
\]
In particular, we have $|\lambda|=0$ and $|x\otimes y|=|x|+|y|$.
A {\em computer} $C$ is a partial-recursive function $C:\s\to\s$, and we denote
the set of all computers by $\Xi$.
Note that our computers do not necessarily have to have prefix-free domain (unless otherwise stated).
If $C\in\Xi$ does not halt on some input $x\in\s$, then we write $C(x)=\infty$ as an abbreviation
for the fact that $C(x)$ is undefined. Thus, we can also interpret computers $C$ as mappings
from $\s$ to $\sbar$, where
\[
   \sbar:=\s\cup \{\infty\}.
\]
As usual, we denote by $K_C(x)$ the {\em Kolmogorov complexity} of the string $x\in\bars$
with respect to the computer $C\in\Xi$
\[
   K_C(x):=\min\left\{|s|\enspace\left\vert\enspace s\in \s, C(s)=x\right.\right\}
\]
or as $\infty$ is this set is empty.

What would be a first, naive try to define algorithmic probability?
Since we do not restrict our approach to prefix computers, we cannot take
Equation~(\ref{EqDefAP}) as a definition. Instead we may try to count how often a string is produced by the computer as output:
\begin{definition}[Output Frequency]
\label{DefAlgFrequency}
\lineclear
For every $C\in\Xi$, $n\in\N_0$ and $s\in\sbar$, we set
\begin{eqnarray*}
\mu_C^{(n)}(s):=\frac{
   \#\{x\in \{0,1\}^n \enspace|\enspace  C(x)=s\}
}{2^n}\,\,.
\end{eqnarray*}
For later use in Section~\ref{SecStationaryComputerProbability}, we also define
for every $C,D\in\Xi$ and $n\in\N_0$
\begin{eqnarray*}
\mu_C^{(n)}(D):=\frac{\#\left\{
x\in \{0,1\}^n \enspace\left\vert\enspace C\emu x D
\right.\right\}
}{2^n}\,\,,
\end{eqnarray*}
where the expression $C\emu{x}D$ is given in Definition~\ref{DefEmulation}.
\end{definition}
Our final definition of algorithmic probability will look very different, but it will
surprisingly turn out to be closely related to this output frequency notion.

The existence of the limit $\lim_{n\to \infty}\mu_C^{(n)}(s)$ depends
on the computer $C$ and may be hard to decide, but
in the special case of prefix computers, the limit exists and agrees with
the classical notion of algorithmic probability as given in Equation~(\ref{EqDefAP}):

\begin{example}[Prefix Computers]
\label{ExPrefix}
A computer $C\in\Xi$ is called {\em prefix} if the following holds:
\[
   C(x)\neq\infty\Longrightarrow C(x\otimes y)=\infty\mbox{ for every }y\neq\lambda.
\]
This means that if $C$ halts on some input $x\in\s$, it must not halt on any
extension $x\otimes y$. Such computers are traditionally studied in algorithmic information
theory. To fit our approach, we need to modify the definition slightly. Call a computer $C_p\in\Xi$ {\em prefix-constant}
if the following holds true:
\[
   C_p(x)\neq \infty \Longrightarrow C_p(x\otimes y)=C_p(x)\mbox{ for every }y\in\s.
\]
It is easy to see that for every prefix computer $C$, one can find a prefix-constant computer $C_p$
with $C_p(x)=C(x)$ whenever $C(x)\neq\infty$. It is constructed in the following way:
Suppose $x\in\s$ is given as input into $C_p$, then it
\begin{itemize}
\item computes the set of all prefixes $\{x_i\}_{i=0}^{|x|}$ of $x$ (e.g. for $x=100$ we have
$x_0=\lambda$, $x_1=1$, $x_2=10$ and $x_3=100$),
\item starts $|x|+1$ simulations of $C$ at the same time, which are supplied with $x_0$
up to $x_{|x|}$ as input,
\item waits until one of the simulations produces an output $s\in\s$ (if this never happens,
$C_p$ will loop forever),
\item finally outputs $s$.
\end{itemize}
Fix an arbitrary string $s\in\bars$. Consider the set
\[
   T^{(n)}(s):=\left\{x\in\s\,\,|\,\, |x|\leq n,C(x)=s\right\}.
\]
Every string $x\in T^{(n)}(s)$ can be extended (by concatenation) to a string $x'$ of
length $n$. By construction, it follows that $C_p(x')=s$. There are $2^{n-|x|}$ possible
extensions $x'$, thus
\[
   \mu_{C_p}^{(n)}(s)=\frac{\sum_{x\in T^{(n)}(s)} 2^{n-|x|}}{2^n}
   =\sum_{x\in\s: |x|\leq n,C(x)=s} 2^{-|x|}\,\,.
\]
It follows that the limit $\mu_{C_p}(s):=\lim_{n\to\infty} \mu_{C_p}^{(n)}(s)$ exists, and it holds
\[
   \mu_{C_p}(s)=\sum_{x\in\s:C(x)=s} 2^{-|x|}\,\,,
\]
so the output frequency as given in Definition~\ref{DefAlgFrequency} converges
for $n\to\infty$ to the classical algorithmic probability as given in Equation~(\ref{EqDefAP}).
Note that $\Omega_C=1-\mu_{C_p}(\infty)$.\qed
\end{example}
It is easy to construct examples of computers which are {\em not} prefix, but which have
an output frequency which either converges, or at least does not tend to zero as $n\to\infty$. Thus,
the notion of output frequency generalizes the idea of algorithmic probability to a larger class of
computers.

\section{Stationary Computer Probability}
\label{SecStationaryComputerProbability}
As explained in the introduction, it will be an essential part of this work to analyze in detail
how ``easily'' one computer $C$ emulates another computer $D$. Our first definition specializes
what we mean by ``emulation'':
\begin{definition}[Emulation]
\label{DefEmulation}
A computer $C\in\Xi$ {\em emulates} the computer $D\in\Xi$ via $x\in\s$, denoted
\[
   C\emu{x}D\qquad \mbox{resp.} \qquad D=\left(C\emu{x}\right),
\]
if $C(x\otimes s)=D(s)$ for every $s\in\s$. We write $C\emu{}D$ if there is
some $x\in\s$ such that $C\emu{x}D$.
\end{definition}
It follows easily from the definition that $C\emu\lambda C$ and
\[
   C\emu{x}D\mbox{ and }D\emu{y}E\Longrightarrow C\emu{x\otimes y}E.
\]

Now that we have defined emulation, it is easy to extend the notion of Kolmogorov
complexity to emulation complexity:
\begin{definition}[Emulation Complexity]
For every $C,D\in\Xi$, the {\em Emulation Complexity} $K_C(D)$
is defined as
\begin{equation}
K_C(D):=\min\left\{|s|\enspace\left\vert\enspace s\in \s,C\emu{s} D\right.\right\}
\end{equation}
or as $\infty$ if the corresponding set is empty.
\end{definition}
Note that similar definitions have already appeared in the literature, see for example
Def. 4.4 and Def. 4.5 in~\cite{Hay}, or the definition of the constant ``${\rm sim}(C)$'' in~\cite{ChaitinBook}.
\begin{definition}[Universal Computer]
Let $\Phi\subset\Xi$ be a set of computers. If there exists a computer $U\in\Phi$ such
that $U\emu{}X$ for every $X\in\Phi$, then $\Phi$ is called {\em connected}, and $U$
is called a {\em $\Phi$-universal computer}. We use the notation $\Phi^U:=\{C\in\Phi\,\,|\,\,
C\mbox{ is }\Phi\mbox{-universal}\}$, and we write
$\overline{\Phi^U}:=\{C\in\Xi\,\,|\,\,C\emu{}D\quad\forall D\in\Phi\mbox{ and }\exists X\in\Phi:
X\emu{}C\}$.
\end{definition}

Note that $\Phi^U\subset\overline{\Phi^U}$ and $\Phi^U=\emptyset\Leftrightarrow\overline{\Phi^U}=\emptyset$.
Examples of connected sets of computers include the set $\Xi$ of all computers
and the set of prefix-constant computers, whereas the set of computers which always halt on every input
cannot be connected, as is easily seen by diagonalization. For convenience, we
give a short proof of the first statement:

\begin{proposition}
The set of all computers $\Xi$ is connected.
\end{proposition}
{\bf Proof.} It is well-known that there is a computer $U$ that takes a description $d_M\in\s$
of any computer $M\in\Xi$ together with some input $x\in\s$ and simulates $M$ on input $x$, i.e.
\[
   U(\langle d_M,x\rangle)=M(x)\mbox{ for every }x\in\s,
\]
where $\langle \cdot,\cdot\rangle:\s\times\s\to\s$ is a bijective and computable encoding
of two strings into one. We can construct the encoding in such a way that
$\langle d_M,x\rangle=\tilde d_M \otimes x$, i.e. the description is encoded into some
prefix code that is appended to the left-hand side of $x$. It follows that
$U\emu{\tilde d_M}M$, and since this works for every $M\in\Xi$, $U$ is $\Xi$-universal.\qed

Here is a basic property of Kolmogorov and emulation complexity:
\begin{theorem}[Invariance of Complexities]
Let $\Phi\subset\Xi$ be connected, then for every $U\in\overline{\Phi^U}$ and $V\in\Phi$, it holds that
\begin{eqnarray*}
   K_U(D)&\leq& K_U(V)+K_V(D)\qquad\mbox{for every }D\in\Phi,\\
   K_U(s)&\leq& K_U(V)+K_V(s)\qquad \mbox{for every }s\in\sbar.
\end{eqnarray*}
\end{theorem}
{\bf Proof.} Since $U\in\overline{\Phi^U}$, it holds $U\emu{}V$. Let $x$ be a shortest string such
that $U(x\otimes t)=V(t)$ for every $t\in\s$, i.e. $|x|=K_U(V)$. If $p_D$ resp. $p_s$
are shortest strings such that $V\emu{p_D}D$ resp. $V(p_s)=s$, then
$|p_D|=K_V(D)$ and $|p_s|=K_V(s)$, and additionally $U\emu{x\otimes p_D}D$ and
$U(x\otimes p_s)=s$. Thus, $K_U(D)\leq |x\otimes p_D|=|x|+|p_D|$ and $K_U(s)\leq |x\otimes p_s|
=|x|+|p_s|$.\qed

Suppose some computer $C\in\Xi$ emulates another computer $E$ via the string
$10$, i.e. $C\emu{10}E$. We can decompose this into two steps: Let $D:=C\emu{1}$,
then
\[
   C\emu{1}D\qquad\mbox{and}\qquad D\emu{0}E.
\]
Similarly, we can decompose every emulation $C\emu{x}D$ into $|x|$ parts,
just by parsing the string $x$ bit by bit, while getting a corresponding
``chain'' of emulated computers. A clear way to illustrate this situation is
in the form of a tree, as shown in Figure~\ref{FigEmulation}. We start at the root
$\lambda$. Since $C\emu{\lambda}C$, this string corresponds to the computer $C$
itself. Then, we are free to choose $0$ or $1$, yielding the computer $\left(C\emu{0}\right)$
or $\left(C\emu{1}\right)=D$ respectively. Ending up with $D$, we can choose the next bit
(taking a $0$ we will end up with $E=\left(D\emu{0}\right)=\left(C\emu{10}\right)$) and so on.

In general, some of the emulated computers
will themselves be elements of $\Phi$ and some not. As in Figure~\ref{FigEmulation},
we can mark every path that leads to a computer that is itself an element of
$\Phi$ by a thick line. (In this case, for example $C,D,E\in\Phi$, but $\left(C\emu{11}\right)\not\in\Phi$.)
If we want to restrict the process of parsing through the tree to the marked (thick)
paths, then we need the following property:

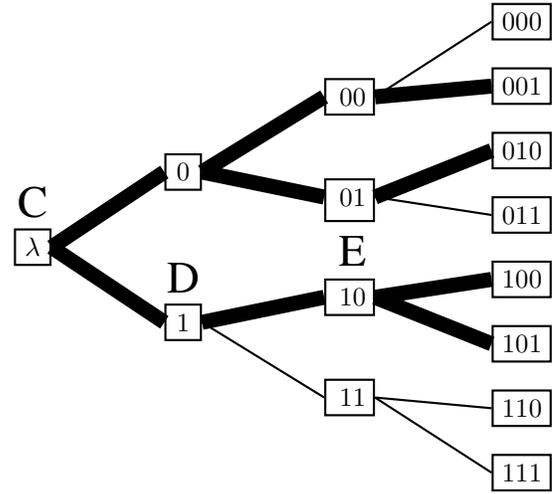
\begin{figure}[!hbt]
\begin{center}
\psset{unit=0.5cm}
\begin{pspicture}(0,-2.5)(14.3,10.5)
  \rput(0.5,4){$ \lambda$}
  \psframe(0,3.5)(1,4.5)
  \rput(4.5,6){$0$}
  \psframe(4,5.5)(5,6.5)
  \rput(4.5,2){$1$}
  \psframe(4,1.5)(5,2.5)
  {\psset{linewidth=0.4}\psline(1,4)(4,6)}
  {\psset{linewidth=0.4}\psline(1,4)(4,2)}
  \rput(9,8){$00$}
  \psframe(8.25,7.5)(9.6,8.5)
  {\psset{linewidth=0.4}\psline(5,6)(8.25,8)}
  \rput(9,2.667){$10$}
  \psframe(8.25,2.167)(9.6,3.167)
  {\psset{linewidth=0.4}\psline(5,2)(8.25,2.667)}
  \rput(9,5.333){$01$}
  \psframe(8.25,4.667)(9.6,5.833)
  {\psset{linewidth=0.4}
  \psline(5,6)(8.25,5.333)}
  \rput(9,0){$11$}
  \psframe(8.25,-0.5)(9.6,0.5)
  \psline(5,2)(8.25,0)
  \rput(13.5,10){$000$}
  \psframe(12.7,9.5)(14.3,10.5)
  \psline(9.6,8)(12.7,10)
  \rput(13.5,-2){$111$}
  \psframe(12.7,-2.5)(14.3,-1.5)
  \psline(9.6,0)(12.7,-2)
  \rput(13.5,-0.2857){$110$}
  \psframe(12.7,-0.7857)(14.3,0.2143)
  \psline(9.6,0)(12.7,-0.2857)
  \rput(13.5,1.4286){$101$}
  \psframe(12.7,0.9286)(14.3,1.9286)
  {\psset{linewidth=0.4}\psline(9.6,2.667)(12.7,1.4286)}
  \rput(13.5,3.1429){$100$}
  \psframe(12.7,2.6429)(14.3,3.6429)
  {\psset{linewidth=0.4}\psline(9.6,2.667)(12.7,3.1429)}
  \rput(13.5,4.857){$011$}
  \psframe(12.7,4.357)(14.3,5.357)
  \psline(9.6,5.333)(12.7,4.857)
  \rput(13.5,6.571){$010$}
  \psframe(12.7,6.071)(14.3,7.071)
  {\psset{linewidth=0.4}\psline(9.6,5.333)(12.7,6.571)}
  \rput(13.5,8.2857){$001$}
  {\psset{linewidth=0.4}\psline(9.6,8)(12.7,8.2857)}
  \psframe(12.7,7.7857)(14.3,8.7857)
  \rput(0.5,5.2){\LARGE C}
  \rput(4.5,3.2){\LARGE D}
  \rput(9,3.9){\LARGE E}
\end{pspicture}
\end{center}
\caption{Emulation as a tree, with a branching subset $\Phi$ (bold lines).}
\label{FigEmulation}
\end{figure}

\begin{definition}
A set of computers $\Phi\subset\Xi$ is called {\em bran\-ching}, if
for every $C\in\Phi$, the following two conditions are satisfied:
\begin{itemize}
\item For every $x,y\in\s$, it holds
\[
   \left(C\emu{x\otimes y}\right)\in\Phi\quad\Longrightarrow\quad \left(C\emu{x}\right)\in\Phi.
\]
\item There is some $x\in\s\setminus\{\lambda\}$ such that $\left(C\emu{x}\right)\in\Phi$.
\end{itemize}
\end{definition}

If $\Phi$ is branching, we can parse through the corresponding marked
subtree without encountering any dead end, with the possibility
to reach every leaf of the subtree.
In particular, these requirements are fulfilled by sets of universal computers:

\begin{proposition}
\label{PropBranching}
Let $\Phi\subset\Xi$ be connected and $\#\overline{\Phi^U}\geq 2$, then $\overline{\Phi^U}$ is branching.
\end{proposition}
{\bf Proof.} Let $C\in\overline{\Phi^U}$ and $\left(C\emu{x\otimes y}\right)\in\overline{\Phi^U}$, then
$\left(C\emu{x\otimes y}\right)\emu{}D$ for every $D\in\Phi$, and so $\left(C\emu{x}\right)\emu{} D$ for
every $D\in\Phi$. Moreover, there is some $X\in\Phi$ such that $X\emu{}C$, so in particular,
$X\emu{}\left(C\emu{x}\right)$. Thus, $\left(C\emu{x}\right)\in\overline{\Phi^U}$.

On the other hand, since $\#\overline{\Phi^U}\geq 2$, there are computers $C,D\in\overline{\Phi^U}$
such that $C\neq D$. By definition of $\overline{\Phi^U}$, there is some $X\in\Phi$ such
that $X\emu{}D$. Since $C$ emulates every computer in $\Phi$, we have $C\emu{}X$, so
$C\emu{z}D$ for some $z\neq\lambda$.
\qed

As illustrated in the bold subtree in Figure~\ref{FigEmulation},
we can define the process of a {\em random walk}
on this subtree if its corresponding computer subset $\Phi$ is branching:
we start at the root $\lambda$, follow the branches, and at every bifurcation,
we turn ``left or right'' (i.e. input an additional $0$ or $1$) with probability $\frac 1 2$.
This random walk generates a probability distribution on the subtree:
\begin{definition}[Path and Computer Probability]
If $\Phi\subset\Xi$ is branching and let $C\in\Phi$, we define the {\em $\Phi$-tree of $C$}
as the set of all inputs $x\in\s$ that make $C$ emulate a computer in $\Phi$ and
denote it by $C^{-1}(\Phi)$, i.e.
\[
   C^{-1}(\Phi):=\left\{x\in\s\,\,|\,\, \left(C\emu{x}\right)\in\Phi\right\}.
\]
To every $x$ in the $\Phi$-tree of $C$, we can associate its {\em path probability}
$\mu_{C^{-1}(\Phi)}(x)$ as the probability of arriving at $x$ on a random walk on
this tree. Formally,
\begin{eqnarray*}
   \mu_{C^{-1}(\Phi)}(\lambda)&:=&1,\\
   \mu_{C^{-1}(\Phi)}(x\otimes b)&:=&\left\{
      \begin{array}{cl}
         \frac 1 2 \mu_{C^{-1}(\Phi)}(x) & \mbox{if }x\otimes\bar b \in C^{-1}(\Phi)\\
         \mu_{C^{-1}(\Phi)}(x) & \mbox{otherwise}
      \end{array}
   \right.
\end{eqnarray*}
for every bit $b\in\{0,1\}$ with $x\otimes b\in C^{-1}(\Phi)$, where $\bar b$ denotes the inverse bit. The associated
$n$-step computer probability of $D\in\Phi$ is defined as the probability of arriving
at computer $D$ on a random walk of $n$ steps on this tree, i.e.
\[
   \mu_C^{(n)}(D|\Phi):=\sum_{x\in\{0,1\}^n:C\emu{x}D} \mu_{C^{-1}(\Phi)}(x).
\]
\end{definition}

It is clear that for $\Phi=\Xi$, we get back the notion of output frequency
as given in Definition~\ref{DefAlgFrequency}: For every $C,D\in\Xi$, it holds
\[
   \mu_C^{(n)}(D)=\mu_C^{(n)}(D|\Xi).
\]
The condition that $\Phi$ shall be branching guarantees that $\sum_{x\in\{0,1\}^n\cap
C^{-1}(\Phi)} \mu_{C^{-1}(\Phi)}(x)=1$ for every $n\in\N_0$, i.e. the conservation
of probability. For example,
the path probability in Figure~\ref{FigEmulation} has values $\mu_{C^{-1}(\Phi)}(0)=
\mu_{C^{-1}(\Phi)}(1)=\frac 1 2$, $\mu_{C^{-1}(\Phi)}(00)=\mu_{C^{-1}(\Phi)}(01)=\frac 1 4$,
$\mu_{C^{-1}(\Phi)}(10)=\frac 1 2$, $\mu_{C^{-1}(\Phi)}(001)=\mu_{C^{-1}(\Phi)}(010)=\frac 1 4
=\mu_{C^{-1}(\Phi)}(100)=\mu_{C^{-1}(\Phi)}(101)$.

It is almost obvious that the random walk on the subtree that generates the
computer probabilities $\mu_C^{(n)}(D|\Phi)$
is a Markov process, which is the statement of the next lemma.
For later reference, we first introduce some notation for the corresponding
transition matrix:

\begin{definition}[Emulation Matrix]
Let $\Phi\subset\Xi$ be branching, and enumerate the computers in $\Phi$ in
arbitrary order: $\Phi=\{C_1,C_2,\ldots\}$. Then, we define the (possibly infinite)
{\em emulation matrix} $E_\Phi$ as
\[
   \left(E_\Phi\right)_{i,j}:=\mu_{C_i}^{(1)}(C_j|\Phi).
\]
\end{definition}

\begin{lemma}[Markovian Emulation Process]
\label{LemChapmanKolm}
If $\Phi\subset\Xi$ is branching, then the computer probabilities $\mu_C^{(n)}(\cdot|\Phi)$
are $n$-step probabilities of some Markov process (which we also denote $\Phi$) whose transition matrix is given
by the emulation matrix $E_\Phi$. Explicitly, with $\delta_i:=(0,\ldots,0,\underbrace{1}_i,0,\ldots)$,
\begin{equation}
   \left(\mu_{C_i}^{(n)}(C_1|\Phi),\mu_{C_i}^{(n)}(C_2|\Phi),\mu_{C_i}^{(n)}(C_3|\Phi),
   \ldots\right)=\delta_i\cdot \left(E_\Phi\right)^n
   \label{eqMarkov}
\end{equation}
for every $n\in\N_0$, and we have the {\em Chapman-Kolmogorov equation}
\begin{equation}
   \mu_C^{(m+n)}(D|\Phi)=\sum_{X\in\Phi}\mu_C^{(m)}(X|\Phi)\mu_X^{(n)}(D|\Phi)
   \label{EqChapmanKolmogorov}
\end{equation}
for every $m,n\in\N_0$. Also, $\Phi\subset\Xi$ (resp. $E_\Phi$) is irreducible if and only if
$C\emu{}D$ for every $C,D\in\Phi$.
\end{lemma}
{\bf Proof.} Equation~(\ref{eqMarkov}) is trivially true for $n=0$ and is shown in
full generality by induction (the proof details are not important for the following
argumentation and can be skipped):
\begin{eqnarray*}
\mu_{C_i}^{(n+1)}(C_j|\Phi)&=&\sum_{\stackrel{x\in\{0,1\}^n,b\in\{0,1\}:}{C_i\emu{x\otimes b}C_j}}
\mu_{C^{-1}(\Phi)}(x\otimes b)\\
&=&\sum_{C_k\in\Phi}\sum_{\stackrel{x\in\{0,1\}^n:}{C_i\emu{x}C_k}}\sum_{\stackrel{b\in\{0,1\}:}
{C_k\emu{b}C_j}} \mu_{C^{-1}(\Phi)}(x\otimes b)\\
&=&\sum_{C_k\in\Phi} \sum_{\stackrel{x\in\{0,1\}^n:}{C_i\emu{x}C_k}}\mu_{C^{-1}(\Phi)}(x)
\mu_{C_k}^{(1)}(C_j|\Phi)\\
&=&\sum_{C_k\in\Phi} \mu_{C_i}^{(n)}(C_k|\Phi)\left(E_\Phi\right)_{k,j}.
\end{eqnarray*}
The Chapman-Kolmogorov equation follows directly from the theory of Markov processes.
The stochastic matrix $E_\Phi$ is irreducible iff for every $i,j\in\N$ there is some $n\in\N$
such that $0<\left((E_\Phi)^n\right)_{i,j}=\mu_{C_i}^{(n)}(C_j|\Phi)$, which is equivalent to
the existence of some $x\in\{0,1\}^n$ such that $C_i\emu{x}C_j$.\qed

The next proposition collects some relations between the emulation Markov process and the corresponding
set of computers. We assume that the reader is familiar with the basic vocabulary from the theory
of Markov chains.
\begin{proposition}[Irreducibility and Aperiodicity]
\label{PropIA}
Let $\Phi\subset\Xi$ be a set of computers.
\begin{itemize}
\item $\Phi$ is irreducible $\Leftrightarrow \Phi=\Phi^U\Leftrightarrow \Phi\subset\overline{\Phi^U}$.
\item If $\Phi$ is connected and $\#\Phi^U\geq 2$, then $\overline{\Phi^U}$ is irreducible
and branching.
\item If $\Phi$ is branching, then we can define the {\em period} of $C\in\Phi$ as
$d(C):={\rm GGT}\left\{n\in\N\,\,\left| \,\,\mu_C^{(n)}(C|\Phi)>0\right.\right\}$
(resp. $\infty$ if this set is empty).
If $\Phi\subset\Xi$ is irreducible, then
$d(C)=d(D)=:d<\infty$ for every $C,D\in\Phi$ holds true. In this case, $d$ will be called the period of $\Phi$,
and if $d=1$, then $\Phi$ is called {\em aperiodic}.
\end{itemize}
\end{proposition}
{\bf Proof.} To prove the first equivalence, suppose that 
$\Phi\subset\Xi$ is irreducible, i.e. for every $C,D\in\Phi$ it holds
$C\emu{}D$. Thus, $\Phi$ is connected and $C\in\Phi^U$, so $\Phi\subset\Phi^U$, and since always
$\Phi^U\subset\Phi$, it follows that $\Phi=\Phi^U$. On the other hand, if $\Phi=\Phi^U$, then for every
$C,D\in\Phi$ it holds $C\emu{}D$, since $C\in\Phi^U$. Thus, $\Phi$ is irreducible.
For the second equivalence, suppose that $\Phi$ is irreducible, thus, $\Phi=\Phi^U\subset\overline{\Phi^U}$.
If on the other hand $\Phi\subset\overline{\Phi^U}$, it follows in particular
for every $C\in\Phi$ that $C\emu{}X$ for every $X\in\Phi$, so $\Phi$ is irreducible.

For the second statement, let $C,X\in\overline{\Phi^U}$ be arbitrary.
By definition of $\overline{\Phi^U}$, it follows that there is some $V\in\Phi$ such
that $V\emu{}X$, and it holds $C\emu{}V$, so $C\emu{}X$, and $\overline{\Phi^U}$ is irreducible.
By Proposition~\ref{PropBranching} and $\#\overline{\Phi^U}\geq\#\Phi^U\geq 2$, $\overline{\Phi^U}$
must be branching.
The third statement is well-known from the theory
of Markov processes. \qed

A basic general result about Markov processes now gives us the desired absolute
computer probability - almost, at least:

\begin{theorem}[Stationary Alg. Computer Probability]
\label{TheStatAlgProp}
Let $\Phi\subset\Xi$ be branching, irreducible and aperiodic. Then, for
every $C,D\in\Phi$, the limit (``computer probability'')
\[
   \mu(D|\Phi):=\lim_{n\to\infty}\mu_C^{(n)}(D|\Phi)
\]
exists and is independent of $C$. There are two possible cases:
\begin{itemize}
\item[(1)] The Markov process which corresponds to $\Phi$ is transient or null recurrent. Then,
\[
   \mu(D|\Phi)=0\mbox{ for every }D\in\Phi.
\]
\item[(2)] The Markov process which corresponds to $\Phi$ is positive recurrent. Then,
\[
   \mu(D|\Phi)>0\mbox{ for every }D\in\Phi,\mbox{ and }\sum_{D\in\Phi}\mu(D|\Phi)=1.
\]
In this case, the vector $\underline{\mu_\Phi}:=\left(\mu(C_1|\Phi),\mu(C_2|\Phi),\ldots\right)$
is the unique stationary probability eigenvector of $E_\Phi$, i.e. the unique
probability vector solution to $\underline{\mu_\Phi}\cdot E_\Phi=\underline{\mu_\Phi}$.
\end{itemize}
\end{theorem}
Note that we have derived this result under quite weak conditions --- e.g. in contrast to
classical algorithmic probability, we do not assume that our computers have prefix-free domain.
Nevertheless, we are left with the problem to determine whether a given set $\Phi$
of computers is positive recurrent (case (2) given above) or not (case (1)).

The most interesting case is $\Phi=\Xi^U$, i.e. the set of computers that are universal in
the sense that they can simulate every other computer without any restriction. This set is
``large'' --- apart from universality, we do not assume any additional property like e.g. being prefix.
By Proposition~\ref{PropIA}, $\Xi^U$ is irreducible and branching.
Moreover, fix any universal computer $U\in\Xi^U$ and
consider the computer $V\in\Xi$, given by
\[
   V(x):=\left\{
      \begin{array}{cl}
         \lambda & \mbox{if }x=\lambda,\\
         V(s) & \mbox{if }x=0\otimes s,\\
         U(s) & \mbox{if }x=1\otimes s.
      \end{array}
   \right.
\]
As $V\emu{1}U$, we know that $V\in\Xi^U$, and since
$V\emu{0}V$, it follows that $\mu_V^{(1)}(V)>0$, and so $d(V)=1=d(\Xi^U)$. Hence
$\Xi^U$ is aperiodic.

So is $\Xi^U$ positive recurrent or not? Unfortunately, the answer turns out to be negative:
$\Xi^U$ is transient. The idea to prove this is to construct a sequence of universal computers $M_1, M_2, M_3,\ldots$
such that each computer $M_i$ emulates the next computer $M_{i+1}$ with large probability,
that is, the probability tends to one as $i$ gets large. Thus, starting the random walk on, say, $M_1$,
it will with positive probability stay on this $M_i$-path forever and never return to any other computer.
See also Figure~\ref{FigVirus} in the Introduction for illustration.
\begin{theorem}[Markoff Chaney Virus]
\label{TheMarkoffChaney}
\lineclear
$\Xi^U$ is transient, i.e. there is no stationary algorithmic computer probability
on the universal computers.
\end{theorem}
{\bf Proof.} Let $U\in\Xi^U$ be an arbitrary universal computer with $U(\lambda)=0$.
We define another computer $M_1\in\Xi$ as follows: If some string $s\in\s$ is supplied
as input, then $M_1$
\begin{itemize}
\item splits the string $s$ into parts $s_1,s_2,\ldots,s_k,s_{tail}$,
such that $s=s_1\otimes s_2 \otimes \ldots\otimes s_k\otimes s_{tail}$
and $|s_i|=i$ for every $1\leq i\leq k$. We also demand that $|s_{tail}|<k+1$ (for example,
if $s=101101101011$, then $s_1=1$, $s_2=01$, $s_3=101$, $s_4=1010$ and $s_{tail}=11$),
\item tests if there is any $i\in\{1,\ldots,k\}$ such that $s_i=0^i$ (i.e. $s_i$
contains only zeros). If yes, then $M_1$ computes and outputs
$U(s_{i+1}\otimes\ldots\otimes s_k\otimes s_{tail})$ (if there are several $i$ with $s_i=0^i$, then
it shall take the smallest one).
If not, then $M_1$ outputs $1^k=\underbrace{1\ldots 1}
_{k\mbox{ times}}$.
\end{itemize}
Let $M_2:=M_1\emu{1}$, $M_3:=M_1\emu{1\otimes 11}$,
$M_4:=M_1\emu{1\otimes 11 \otimes 111}$ and so on,
in general $M_n:=M_1\emu{1^{1+2+\ldots+(n-1)}}$ resp. $M_i\emu{1^i}M_{i+1}$.
We also have $M_i\emu{0^i}U$, so $M_i\in\Xi^U$ for every $i\in\N$. Thus, the computers
$M_i$ are all universal. Also, since $M_i(\lambda)=M_1(1^{1+\ldots+(i-1)})=1^{i-1}$, the computers
$M_i$ are mutually different from each other, i.e. $M_i\neq M_j$ for $i\neq j$.
Now consider the computers $M_i\emu{s}$ for $|s|=i$, but $s\neq 0^i$. It holds
$M_i(s\otimes x)=M_1(1\otimes 11\otimes\ldots\otimes 1^{i-1}\otimes s \otimes x)$.
The only property of $s$ that affects the outcome of $M_1$'s computation is the property
to be different from $0^i$. But this property is shared by the string $1^i$, i.e.
$M_1(1\otimes  11 \otimes\ldots\otimes 1^{i-1}\otimes s\otimes x)=M_1(1\otimes 11 \otimes \ldots
\otimes 1^{i-1}\otimes 1^i\otimes x)$, resp. $M_i(s\otimes x)=M_i(1^i\otimes x)$ for every $x\in\s$.
Thus, $\left(M_i\emu{s}\right)=\left(M_i\emu{1^i}\right)=M_{i+1}$ for every $0^i\neq s\in\{0,1\}^i$, and so
\[
   \mu_{M_i}^{(i)}(M_{i+1}|\Xi^U)=1-2^{-i}\mbox{ for every }i\in\N.
\]
Iterated application of the Chapman-Kolmogorov equation (\ref{EqChapmanKolmogorov}) yields for every $n\in\N$
\begin{eqnarray*}
   \mu_{M_1}^{(1+2+\ldots+(n-1))}(M_n|\Xi^U)&\geq&
   \mu_{M_1}^{(1)}(M_2|\Xi^U)\cdot\mu_{M_2}^{(2)}(M_3|\Xi^U)\\
   &&\ldots\cdot \mu_{M_{n-1}}^{(n-1)}(M_n|\Xi^U)\\
   &=&\prod_{i=1}^{n-1} \left(1-2^{-i}\right)\\
   &>&\prod_{i=1}^\infty \left(1-2^{-i}\right)=0.2887\ldots
\end{eqnarray*}
With at least this probability, the Markov process corresponding to $\Phi$ will
follow the sequence of computers $\{M_i\}_{i\in\N}$ forever, without ever returning
to $M_1$. (Note that also the intermediately emulated computers like $M_1\emu{11}$
are different from $M_1$, since $M_1(\lambda)=\lambda$, but $\left(M_1\emu{11}\right)(\lambda)\neq\lambda$.)
Thus, the eventual return probability to $M_1$ is strictly less than $1$.\qed

In this proof, every computer $M_{i+1}$ is a modified copy of its ancestor $M_i$.
In some sense, $M_1$ can be seen as some kind of ``computer virus'' that undermines the
existence of a stationary computer probability. The theorem's name ``Markoff Chaney Virus'' was inspired by
a fictitious character in Robert Anton Wilson's ``Illuminatus!'' trilogy\footnote{``{\em The Midget, whose
name was Markoff Chaney, was no relative of the famous Chaneys of Hollywood, but people did keep making
jokes about that. [...] Damn the science of mathematics itself, the line, the square, the average,
the whole measurable world that pronounced him a bizarre random factor. Once and for all, beyond fantasy,
in the depth of his soul he declared war on the ``statutory ape'', on law and order, on predictability,
on negative entropy. He would be a random factor in every equation; from this day forward, unto death,
it would be civil war: the Midget versus the Digits...
}''}.

The set $\Xi^U$ is in some sense too large to allow the existence of stationary algorithmic
probability distribution. Yet, there exist computer sets $\Phi$ that are actually positive recurrent
and thus have such a probability distribution; here is an explicit example:
\begin{example}[A Positive Recurrent Computer Set]
\label{ExPosRecurrent}
Fix an arbitrary string $u\in\s$ with $|u|\geq 2$, and let $U$ be a universal computer, i.e. $U\in\Xi^U$,
with the property that it emulates every other computer via some string that does not contain $u$ as a substring,
i.e.
\[
   \forall D\in\Xi \enspace\exists d\in\s:U\emu{d}D\mbox{ and }u\mbox{ not substring of }d.
\]
If $C\in\Xi$ is any computer, define a corresponding computer $C_{u,U}$ by
$C_{u,U}(x)=U(y)$ if $x=w\otimes u \otimes y$ and $y$ does not contain $u$ as a substring, and as
$C_{u,U}(x)=C(x)$ otherwise (that is, if $x$ does not contain $u$). The string $u$ is a ``synchronizing word'' for
the computer $C_{u,U}$, in the sense that any occurrence of $u$ in the input forces $C_{u,U}$ to ``reset'' and
to emulate $U$.

We get a set of computers
\[
   \Phi_{u,U}:=\{C_{u,U}\,\,|\,\, C\in\Xi\}.
\]
Whenever $x$ does not contain $u$ as a substring, it holds
\[
   C\emu x D \enspace\Rightarrow\enspace  C_{u,U} \emu x  D_{u,U}.
\]
It follows that $V:=U_{u,U}$ is a universal computer for $\Phi_{u,U}$. Thus $\Phi_{u,U}$ is connected,
and it is easy to see that $\Phi_{u,U}^U=\overline{\Phi_{u,U}^U}$ and $\# \Phi_{u,U}^U \geq 2$.
According to to Proposition~\ref{PropIA},
$\Phi_{u,U}^U$ is irreducible and branching. An argument similar to that before Theorem~\ref{TheMarkoffChaney} (where it
was proved that $\Xi^U$ is aperiodic) proves that $\Phi_{u,U}^U$ is also aperiodic.
Moreover, by construction it holds for every computer $C\in\Phi_{u,U}^U$ and $\ell:=|u|$
\[
   \mu_C^{(\ell)}(V|\Phi_{u,U}^U)\geq 2^{-\ell}.
\]
The Chapman-Kolmogorov equation~(\ref{EqChapmanKolmogorov}) then yields
\begin{eqnarray*}
\mu_C^{(n+\ell)}(V|\Phi_{u,U}^U)&=& \sum_{X\in\Phi_{u,U}^U} \mu_C^{(n)}(X|\Phi_{u,U}^U)\mu_X^{(\ell)}(V|\Phi_{u,U}^U)\\
&\geq& 2^{-\ell}\sum_{X\in\Phi_{u,U}^U} \mu_C^{(n)}(X|\Phi_{u,U}^U)=2^{-\ell}.
\end{eqnarray*}
Consequently, $\limsup_{n\to\infty} \mu_C^{(n)}(V|\Phi_{u,U}^U)\geq 2^{-\ell}$. According to Theorem~\ref{TheStatAlgProp},
it follows that $\Phi_{u,U}^U$ is positive recurrent. In particular, $\mu(V|\Phi_{u,U}^U)\geq 2^{-|u|}$.
Note also that $\# \Phi_{u,U}^U=\infty$, so we do not have the trivial situation of a finite computer set.
\end{example}

Obviously, the computer set $\Phi_{u,U}$ in the previous example depends on the choice of the string $u$ and the computer $U$;
different choices yield different computer sets and different probabilities.
In the next section, we will see in Theorem~\ref{TheNonUniqueness}
that every positive recurrent computer set contains an unavoidable ``amount of arbitrariness'', and this fact has an interesting physical
interpretation.

Given any positive recurrent computer set $\Phi$ (as in the previous example), the actual numerical values of the
corresponding stationary computer probability $\mu(\cdot|\Phi)$ will in general be
noncomputable. For this
reason, the following lemma may be interesting, giving a rough bound on stationary
computer probability in terms of emulation complexity:
\begin{lemma}Let $\Phi\subset\Xi$ be positive recurrent\footnote{In the following,
by stating that some computer set $\Phi\subset\Xi$ is positive recurrent, we shall always
assume that $\Phi$ is also branching, irreducible and aperiodic.
}. Then, for every $C,D\in\Phi$,
we have the inequality
\[
   2^{-K_C(D)}\leq \frac{\mu(D|\Phi)}{\mu(C|\Phi)} \leq 2^{K_D(C)}.
\]
\end{lemma}
{\bf Proof.} We start with the limit $m\to\infty$ in the Chapman-Kolmogorov equation
(\ref{EqChapmanKolmogorov}) and obtain
\begin{eqnarray*}
\mu(D|\Phi)&=&\sum_{U\in\Phi} \mu(U|\Phi)\mu_U^{(n)}(D|\Phi)\\
&\geq& \mu(C|\Phi)\mu_C^{(n)}(D|\Phi)
\end{eqnarray*}
for every $n\in\N_0$. Next, we specialize $n:=K_C(D)$, then $\mu_C^{(n)}(D|\Phi)\geq 2^{-n}$.
This proves the left hand side of the inequality. The right hand side can be obtained
simply by interchanging $C$ and $D$.\qed

\section{Symmetries and String Probability}
\label{SecSymmetries}
The aim of this section is twofold: on the one hand, we will derive an alternative
proof of the non-existence of a stationary computer probability distribution on $\Xi^U$ (which
we have already proved in Theorem~\ref{TheMarkoffChaney}). The benefit of this alternative
proof will be to generalize our no-go result much further: it will supply us with an interesting
physical interpretation why getting rid of machine-dependence must be impossible. We discuss
this in more detail in Section~\ref{SecConclusions}.

On the other hand, we would like to explore
what happens for computer sets $\Phi$ that actually {\em are} positive recurrent. In particular,
we show that such sets generate a natural algorithmic probability on the {\em strings} --- after all,
finding such a probability distribution was our aim from the beginning (cf. the Introduction).
Actually, this string probability turns out to be useful in proving our no-go generalization.
Moreover, it shows that the hard part is really to define computer probability --- once this is
achieved, string probability follows almost trivially.

Here is how we define string probability. While computer probability $\mu(C|\Phi)$
was defined as the probability
of encountering $C$ on a random walk on the $\Phi$-tree, we analogously define
the probability of a string $s$ as the probability of getting the
output $s$ on this random walk:

\begin{definition}[String Probability]
\label{DefStringProbability}
Let $\Phi\subset\Xi$ be branching and let $C\in\Phi$. The $n$-step
string probability of $s\in\sbar$ is defined as the probability of arriving
at output $s$ on a random walk of $n$ steps on the $\Phi$-tree of $C$, i.e.
\[
   \mu_C^{(n)}(s|\Phi):=\sum_{x\in\{0,1\}^n\cap C^{-1}(\Phi):C(x)=s}\mu_{C^{-1}(\Phi)}(x).
\]
\end{definition}

\begin{theorem}[Stationary Algorithmic String Probability]
\label{TheStatAlgSP}
If $\Phi\subset\Xi$ is positive recurrent, then for every $C\in\Phi$ and $s\in\bars$
the limit
\begin{eqnarray*}
   \mu(s|\Phi)&:=&\lim_{n\to\infty}\mu_C^{(n)}(s|\Phi)\\
   &=&\sum_{U\in\Phi} \mu(U|\Phi)\mu_U^{(0)}(s|\Phi)=\sum_{U\in\Phi:U(\lambda)=s}\mu(U|\Phi)
\end{eqnarray*}
exists and is independent of $C$.
\end{theorem}
{\bf Proof.} It is easy to see from the definition of $n$-step string probability
that
\[
   \mu_C^{(n)}(s|\Phi)=\sum_{U\in\Phi:U(\lambda)=s}\mu_C^{(n)}(U|\Phi).
\]
Taking the limit $n\to\infty$, Theorem~\ref{TheStatAlgProp} yields equality
of left and right hand side, and thus existence of the limit and independence
of $C$.\qed

In general, $\mu(\cdot|\Phi)$ is a probability distribution on $\sbar$ rather than on $\s$,
i.e. the undefined string can have positive probability, $\mu(\infty|\Phi)>0$, so
$\sum_{s\in\s}\mu(s|\Phi)<1$.

We continue by showing a Chapman-Kolmogorov-like equation (analogous to Equation~(\ref{EqChapmanKolmogorov}))
for the string probability. Note that this equation differs from the much deeper result of Theorem~\ref{TheEqDef}
in the following sense: it describes a weighted average of probabilities $\mu_U^{(n)}(s|\Phi)$, and
those probabilities do not only depend on the computer $U$ (as in Theorem~\ref{TheEqDef}), but also on
the choice of the subset $\Phi$.
\begin{proposition}[Chapman-Kolmogorov for String Prob.]
\label{EqCKString}
If $\Phi\subset\Xi$ is positive recurrent, then
\[
   \mu_C^{(m+n)}(s|\Phi)=\sum_{U\in\Phi}\mu_C^{(m)}(U|\Phi)\mu_U^{(n)}(s|\Phi)
\]
for every $C\in\Phi$, $m,n\in\N_0$ and $s\in\sbar$.
\end{proposition}
{\bf Proof.} For $x,y\in\sbar$, we use the notation
\[
   \delta_{x,y}:=\left\{
      \begin{array}{cl}
         0 & \mbox{if }x\neq y\\
         1 & \mbox{if }x=y
      \end{array}
   \right.
\]
and calculate
\begin{eqnarray*}
\mu_C^{(m+n)}(s|\Phi)&=&\sum_{x\in\{0,1\}^{m+n}\cap C^{-1}(\Phi)} \mu_{C^{-1}(\Phi)}(x)\cdot \delta_{s,C(x)}\\
&=&\sum_{U\in\Phi}\mu_C^{(m+n)}(U|\Phi)\mu_U^{(0)}(s|\Phi)\\
&=&\sum_{U\in\Phi}\sum_{V\in\Phi}\mu_C^{(m)}(V|\Phi)\mu_V^{(n)}(U|\Phi)\mu_U^{(0)}(s|\Phi)\\
&=&\sum_{V\in\Phi}\mu_C^{(m)}(V|\Phi)\sum_{U\in\Phi}\mu_V^{(n)}(U|\Phi)\mu_U^{(0)}(s|\Phi).
\end{eqnarray*}
The second sum equals $\mu_V^{(n)}(s|\Phi)$ and the claim follows.\qed

For prefix computers $C$,
algorithmic probability $P_C(s)$ of any string $s$ as defined in Equation~(\ref{EqDefAP}) and
the expression $2^{-K_C(s)}$ differ only by a multiplicative constant~\cite{LiVitanyi}. Here is
an analogous inequality for stationary string probability:

\begin{lemma}Let $\Phi\subset\Xi$ be positive recurrent and $C\in\Phi$ some
arbitrary computer, then
\[
   \mu(s|\Phi)\geq \mu(C|\Phi)\cdot 2^{-K_C(s)}\qquad\mbox{for all }s\in\s.
\]
\end{lemma}
{\bf Proof.} We start with the limit $m\to\infty$ in the Chapman-Kolmogorov
equation given in Proposition~\ref{EqCKString} and get
\begin{eqnarray*}
   \mu(s|\Phi)&=&\sum_{U\in\Phi}\mu(U|\Phi)\mu_U^{(n)}(s|\Phi)\\
   &\geq&\mu(C|\Phi)\mu_C^{(n)}(s|\Phi)
\end{eqnarray*}
for every $n\in\N_0$. Then we specialize $n:=K_C(s)$ and use $\mu_C^{(n)}(s|\Phi)\geq 2^{-n}$
for this choice of $n$.\qed

Looking for further properties of stationary string probability, it seems reasonable
to conjecture that, for many computer sets $\Phi$, a string $s\in\s$ (like $s=10111$)
and its inverse $\bar s$ (in this case
$\bar s=01000$) have the same probability $\mu(s|\Phi)=\mu(\bar s|\Phi)$, since both seem to be in some sense
algorithmically equivalent. A general approach to prove such conjectures is to
study {\em output transformations}:

\begin{definition}[Output Transformation $\sigma$]
\lineclear
Let $\sigma:\s\to\s$ be a computable permutation. For every $C\in\Xi$, the map $\sigma\circ C$ is
itself a computer, defined by $\sigma\circ C(x):=\sigma(C(x))$. The map $C\mapsto \sigma\circ C$ will
be called an {\em output transformation} and will also be denoted $\sigma$.
Moreover, for computer sets $\Phi\subset\Xi$, we use the notation
\[
   \sigma\circ\Phi:=\left\{\sigma\circ C\,\,|\,\,C\in\Phi\right\}.
\]
\end{definition}
Under reasonable conditions, string and computer probability are invariant with respect to output transformations:
\begin{theorem}[Output Symmetry]
\label{TheOutputSym}
Let $\Phi\subset\Xi$ be positive recurrent and closed\footnote{We say that a computer set $\Phi\subset\Xi$
is {\em closed} with respect to some transformation $T:\Xi\to\Xi$ if $\Phi\supset T(\Phi):=\left\{
T(C)\,\,|\,\,C\in\Phi\right\}$.
} with respect to some output transformation $\sigma$ and its inverse $\sigma^{-1}$. Then,
we have for every $C\in\Phi$
\[
   \mu(C|\Phi)=\mu(\sigma\circ C|\Phi)
\]
and for every $s\in\s$
\[
   \mu(s|\Phi)=\mu(\sigma(s)|\Phi).
\]
\end{theorem}
{\bf Proof.} Note that $\Phi=\sigma\circ\Phi$.
Let $C,D\in\Phi$. Suppose that $C\emu{b}D$ for some bit $b\in\{0,1\}$. Then,
\begin{eqnarray*}
\sigma\circ C(b\otimes x) = \sigma(D(x))=\sigma\circ D(x).
\end{eqnarray*}
Thus, we have $\sigma\circ C\emu{b}\sigma\circ D$. It follows
for the $1$-step transition probabilities that
\begin{eqnarray*}
\left(E_\Phi\right)_{i,j}&=&\mu_{C_i}^{(1)}(C_j|\Phi)\\
&=&\mu_{\sigma\circ C_i}^{(1)}(\sigma\circ C_j|\Phi)=\left(E_{\sigma\circ\Phi}\right)_{i,j}
\end{eqnarray*}
for every $i,j$. Thus, the emulation matrix $E_\Phi$ does not change if every computer $C$ (or rather its
number in the list of all computers) is exchanged with (the number of) its transformed computer
$\sigma\circ C$ yielding the transformed emulation matrix $E_{\sigma\circ\Phi}$.
But then, $E_\Phi$ and $E_{\sigma\circ\Phi}$ must have the same unique stationary probability
eigenvector
\[
   \underline{\mu_\Phi}=\left(\mu(C_k|\Phi)\right)_{k=1}^{\#\Phi}=
   \underline{\mu_{\sigma\circ\Phi}}
   =\left(\mu(\sigma\circ C_k|\Phi)\right)_{k=1}^{\#\Phi}.
\]
This proves the first identity, while the second identity follows
from the calculation
\begin{eqnarray*}
\mu(s|\Phi)&=&\sum_{U\in\Phi:U(\lambda)=s}\mu(U|\Phi)=\sum_{U\in\Phi:U(\lambda)=s}
\mu(\sigma\circ U|\Phi)\\
&=&\sum_{V\in\sigma\circ\Phi:V(\lambda)=\sigma(s)}\mu(V|\Phi)=\mu(\sigma(s)|\Phi).
\qquad\enspace\,\,\mbox{\qed}
\end{eqnarray*}
Thus, if some computer set $\Phi\subset\Xi$ contains e.g. for every computer $C$
also the computer $\bar C$ which always outputs the bitwise inverse of $C$, then
$\mu(s|\Phi)=\mu(\bar s|\Phi)$ holds. In some sense, this shows that the approach
taken in this paper successfully eliminates
properties of single computers (e.g. to prefer the string $10111$ over $01000$)
and leaves only general algorithmic properties related to the set of computers.

Moreover, Theorem~\ref{TheOutputSym} allows for an alternative proof that $\Xi^U$ and
similar computer sets cannot be positive recurrent. We call a set of computable permutations
$S:=\{\sigma_i\}_{i\in\N}$ {\em cyclic} if every string $s\in\s$ is mapped to
infinitely many other strings by application of finite compositions of those permutations, i.e. if
for every $s\in\s$
\[
   \#\left\{\left. \sigma_{i_1}\circ \sigma_{i_2}\circ\ldots\circ\sigma_{i_N}(s)\,\,\right|\,\,N\in\N,\enspace
   i_n\in\N\right\}=\infty,
\]
and if $S$ contains with each permutation $\sigma$ also its inverse $\sigma^{-1}$. Then, many
computer subset cannot be positive recurrent:
\begin{theorem}[Output Symmetry and Positive Recurrence]
\label{ThePosRecSym}
Let $\Phi\subset\Xi$ be closed with respect to a cyclic set of output transformations,
then $\Phi$ is not positive recurrent.
\end{theorem}
{\bf Proof.} Suppose $\Phi$ is positive recurrent. Let $S:=\{\sigma_i\}_{i\in\N}$
be the corresponding cyclic set of output transformations. Let $s\in\s$ be an arbitrary string,
then for every composition $\sigma:=\sigma_{i_1}\circ\ldots\circ\sigma_{i_N}$, 
we have by Theorem~\ref{TheOutputSym}
\[
   \mu(s|\Phi)=\mu(\sigma(s)|\Phi).
\]
Since $S$ is cyclic, there are infinitely many such transformations $\sigma$, producing
infinitely many strings $\sigma(s)$ which all have the same probability. It follows
that $\mu(s|\Phi)=0$. Since $s\in\s$ was arbitrary, this is a contradiction.\qed

Again, we conclude that $\Xi^U$ is not positive recurrent, since this computer set
is closed with respect to {\em all} output transformations.

Although $\Xi^U$ is not positive recurrent, there might be a unique, natural, ``maximal'' or ``most interesting''
subset $\Phi\subset\Xi$ which is positive recurrent. What can we say about this idea? In fact, the
following theorem says that this is also impossible. As this theorem is only a simple generalization of Theorem~\ref{TheOutputSym},
we omit the proof.
\begin{theorem}[Non-Uniqueness]
\label{TheNonUniqueness}
If $\Phi\subset\Xi$ is positive recurrent, then so is $\sigma\circ\Phi$ for
every computable permutation (=output transformation) $\sigma$. Moreover,
\[
   \mu(C|\Phi)=\mu(\sigma\circ C|\sigma\circ\Phi)
\]
for every $C\in\Phi$, and
\[
   \mu(s|\Phi)=\mu(\sigma(s)|\sigma\circ\Phi)
\]
for every $s\in\sbar$.
\end{theorem}
This means that there cannot be a unique ``natural'' positive recurrent computer set $\Phi$:
for every such set $\Phi$, there exist output transformations $\sigma$
such that $\sigma\circ\Phi\neq\Phi$ (this follows from
Theorem~\ref{ThePosRecSym}). But then, Theorem~\ref{TheNonUniqueness} proves that
$\sigma\circ\Phi$ is positive recurrent, too --- and it is thus another candidate
for the ``most natural'' computer set.

\section{Conclusions and Interpretation}
\label{SecConclusions}
We have studied a natural approach to get rid of machine-dependence in the definition of algorithmic
probability. The idea was to look at a Markov process of universal computers emulating each other,
and to take the stationary distribution as a natural probability measure on the computers.

This approach was only partially successful: as the corresponding Markov process on the set of {\em all}
computers is not positive recurrent and thus has no unique stationary distribution, one has to choose a subset
$\Phi$ of the computers, which introduces yet another source of ambiguity.

However, we have shown (cf. Example~\ref{ExPosRecurrent}) that there exist non-trivial, infinite sets $\Phi$ of
computers that are actually positive recurrent and possess a stationary algorithmic probability distribution.
This distribution has beautiful properties and eliminates at least some of the machine-dependence arising from
choosing a single, arbitrary universal computer as a reference machine (e.g. Theorem~\ref{TheOutputSym}). It gives probabilities for computers as well
as for strings (Theorem~\ref{TheStatAlgSP}), agrees with the average output frequency (Theorem~\ref{TheEqDef}),
and does not assume that the computers have any specific
structural property like e.g. being prefix-free.

The second main result can be stated as follows: {\em There is no way to get completely rid of machine-dependence,}
neither in the approach of this paper nor in any other similar but different approach. To understand why this is true,
recall that the main reason for our no-go result was the symmetry of computer probability with respect to {\em output
transformations} $C\mapsto\sigma\circ C$, where $\sigma$ is a computable permutation
on the strings. This can be seen in two places:
\begin{itemize}
\item In Theorem~\ref{ThePosRecSym}, this symmetry yields the result that any computer set which is ``too large'' (like $\Xi^U$)
cannot be positive recurrent.
\item Theorem~\ref{TheNonUniqueness} states that if a set $\Phi$ is positive recurrent, then $\sigma\circ\Phi$
must be positive recurrent, too. Since in this case $\Phi\neq\sigma\circ\Phi$ for many $\sigma$, this means
that there cannot be a unique ``natural'' choice of the computer set $\Phi$.
\end{itemize}
Output transformations have a natural physical interpretation as ``renaming the objects that the strings are describing''.
To see this, suppose we want to define the complexity of the microstate of a box of gas in thermodynamics (this
can sometimes be useful, see~\cite{LiVitanyi}). Furthermore, suppose we are only interested in a coarse-grained description
such that there are only countably many possibilities what the positions, velocities etc. of the gas particles might look like.
Then, we can encode every microstate into a binary string, and define the complexity of a microstate as the complexity
of the corresponding string (assuming that we have fixed an arbitrary complexity measure $K$ on the strings).

But there are always many different possibilities how to encode the microstate into a string (specifying
the velocities in different data formats, specifying first the positions and then the velocities or the other way round etc.).
If every encoding is supposed to be one-to-one and can be achieved by some machine, then two different encodings
will always be related to each other by a computable permutation.

In more detail, if one encoding $e_1$ maps microstates $m$ to encoded strings $e_1(m)\in\s$, then another encoding $e_2$
will map microstates $m$ to $e_2(m)=\sigma(e_1(m))$, where $\sigma$ is a computable permutation on the strings
(that depends on $e_1$ and $e_2$). Choosing encoding $e_1$, a microstate $m$ will be assigned the complexity $K(e_1(m))$,
while for encoding $e_2$, it will be assigned the complexity $K(\sigma\circ e_1(m))$. That is, there is an unavoidable ambiguity
which arises from the arbitrary choice of an encoding scheme. Switching between the two encodings amounts to ``renaming''
the microstates, and this is exactly an output transformation in the sense of this paper.

Even if we do not have the situation that the strings shall describe physical objects, we encounter a similar
ambiguity already in the definition of a computer: a computer, i.e. a partial recursive function, is described
by a Turing machine computing that function. Whenever we look at the output of a Turing machine, we have to ``read'' the output
from the machine's tape which can potentially be done in several inequivalent ways, comparable to the different
``encodings'' described above.

Every kind of attempt to get rid of those additive constants in Kolmogorov complexity will have to face
this ambiguity of ``renaming''. This is why we think that all those attempts must fail.

\appendix[String Probability is the Weighted Average of Output Frequency]
This appendix is rather technical and can be skipped on first reading.
Its aim is to prove Theorem~\ref{TheEqDef}. This theorem says that the string probability which
has been introduced in Definition~\ref{DefStringProbability} in Section~\ref{SecSymmetries} is exactly what
we really wanted to have from the beginning: in the introduction, our main motivation to find a probability measure
on the computers was to define machine-independent algorithmic probability of strings as the weighted mean over
all universal computers as stated in Equation~(\ref{EqProbIdea}). Theorem~\ref{TheEqDef} says that string probability
can be written exactly in this way, given some natural assumptions on the reference set of computers.

Note that Theorem~\ref{TheEqDef} is a surprising result for the following reason: string probability, as defined
in Definition~\ref{DefStringProbability}, only depends on the outputs of the computers on the ``universal subtree'',
that is, on the leaves in Figure~\ref{FigEmulation} which correspond to bold lines. But output frequency, as given
on the right-hand side in Theorem~\ref{TheEqDef} and defined in Definition~\ref{DefAlgFrequency}, counts the
outputs on {\em all} leaves --- that is, output frequency
is a property of a single computer, not of the computer subset that is underlying the emulation Markov process.

In Section~\ref{SecSymmetries}, we have studied output transformations on computers --- the key idea in this appendix
will be to study {\em input transformations} instead. So what is an input transformation?
If $\sigma:\s\to\s$ is a computable permutation on the strings
and $C\in\Xi$ is some computer, we might consider the transformed computer $C\circ\sigma$,
given by $(C\circ\sigma)(s):=C(\sigma(s))$. But this turns out not to be useful, since such transformations
do not preserve the emulation structure. In fact, the most important and useful property of output
transformations in Section~\ref{SecSymmetries} was that they preserve the emulation structure: it holds
\[
   C\emu{s}D\quad\Longleftrightarrow\quad\sigma\circ C\emu{s}\sigma\circ D.
\]
But for transformations like $C\mapsto C\circ\sigma$, there is no such identity --- hence we have to look for a different approach.
It turns out that a successful approach is to look only at a restricted class of permutations, and also to introduce
equivalence classes of computers:
\begin{definition}[Equivalence Classes of Computers]
\lineclear
For every $k\in\N$, two computers $C,D\in\Xi$ are called $k$-equivalent, denoted
$C\stackrel{k}{\sim} D$, if $C(x)=D(x)$ for every $x\in\s$ with $|x|\geq k$.
We denote the corresponding equivalence classes by $[C]_k$ and set
\[
   [\Phi]_k:=\{|C]_k\,\,|\,\,C\in\Phi\}.
\]
A computer set $\Phi\subset\Xi$ is called {\em complete} if for every $C\in\Phi$
and $k\in\N$ it holds $[C]_k\subset\Phi$. If $\Phi\subset\Xi$ is positive recurrent
and complete, we set for every $[C]_k\in[\Phi]_k$
\[
   \mu([C]_k|\Phi):=\sum_{C\in[C]_k}\mu(C|\Phi).
\]
It is easy to see that for every $C,D\in\Xi$ it holds
\[
   C\stackrel k \sim D\Leftrightarrow \left[\left(C\emu s\right)\stackrel k \sim \left(D\emu s\right)
   \mbox{ for every }s\in\s\right],
\]
thus, the definition $\mu_{[C]_k}^{(n)}([D]_k|\Phi):=\sum_{D\in[D]_k}\mu_C^{(n)}(D|\Phi)$
makes sense for $n\in\N$ and $[C]_k,[D]_k\in[\Phi]_k$
and is independent of the choice of the representative $C\in[C]_k$.
Enumerating the equivalence classes $[\Phi]_k=\left\{[C_1]_k,[C_2]_k,[C_3]_k,\ldots\right\}$
in arbitrary order, we can define an associated emulation matrix $\mathcal{E}_{\Phi,k}$ as
\[
   \left(\mathcal{E}_{\Phi,k}\right)_{i,j}:=\mu_{[C_i]_k}^{(1)}([C_j]_k|\Phi).
\]
\end{definition}

It is easily checked that if $\Phi$ is positive recurrent, then the Markov process described
by the transition matrix $\mathcal{E}_{\Phi,k}$ must also be irreducible, aperiodic
and positive recurrent, and $\underline{\mu_{\Phi,k}}:=\left(
\mu([C_1]_k|\Phi),\mu([C_2]_k|\Phi),\mu([C_3]_k|\Phi),\ldots\right)$ is the unique
probability vector solution to the equation $\underline{\mu_{\Phi,k}}\cdot
\mathcal{E}_{\Phi,k}=\underline{\mu_{\Phi,k}}$.

Now we can define input transformations:
\begin{definition}[Input Transformation $\mathcal{I}_\sigma$]
\label{DefInputTrafo}
\lineclear
Let $\sigma:\{0,1\}^n\to\{0,1\}^n$ be a permutation such that
there is at least one string $x\in\{0,1\}^n$ for which $x_1\neq \sigma(x)_1$,
where $x_1$ denotes the first bit of $x$.
For every $s\in\s$, let $\mathcal{I}_\sigma(s)$ be the string that is generated
by applying $\sigma$ to the last $n$ bits of $s$ (e.g. if $n=1$, $\sigma(1)=0$ and $s=1011$,
then $\mathcal{I}_\sigma(s)=1010$). If $|s|<n$, then $\mathcal{I}_\sigma(s):=s$.
For every $C\in\Xi$, the $\mathcal{I}_\sigma$-transformed computer $\mathcal{I}_\sigma(C)$ is defined by
\[
   \left(\mathcal{I}_\sigma(C)\right)(s):=C(\mathcal{I}_\sigma(s))\mbox{ for every }s\in\s.
\]
We call $|\sigma|:=n$ the {\em order} of $\sigma$.
Moreover, we use the notation
\[
   \mathcal{I}_\sigma(\Phi):=\left\{\mathcal{I}_\sigma(C)\,\,|\,\,C\in\Phi\right\}.
\]
\end{definition}
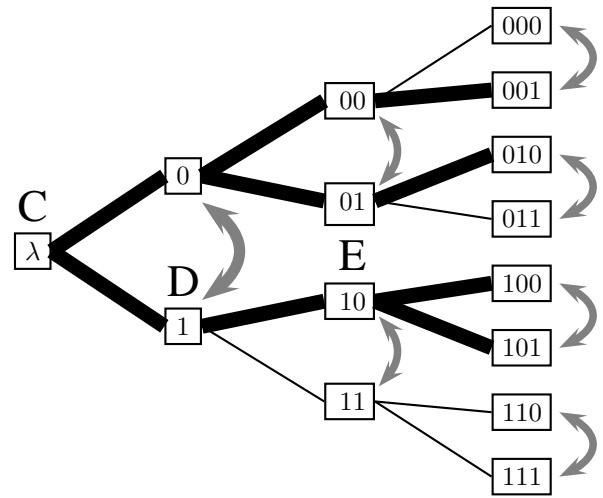
\begin{figure}[!hbt]
\begin{center}
\psset{unit=0.5cm}
\begin{pspicture}(0,-2)(14.3,10.5)
  \rput(0.5,4){$ \lambda$}
  \psframe(0,3.5)(1,4.5)
  \rput(4.5,6){$0$}
  \psframe(4,5.5)(5,6.5)
  \rput(4.5,2){$1$}
  \psframe(4,1.5)(5,2.5)
  {\psset{linewidth=0.4}\psline(1,4)(4,6)}
  {\psset{linewidth=0.4}\psline(1,4)(4,2)}
  \rput(0,0){}
  {\psset{linewidth=0.3,linecolor=gray}\pscurve{<->}(5,5.3)(6,4)(5,2.7)}
  \rput(9,8){$00$}
  \psframe(8.25,7.5)(9.6,8.5)
  {\psset{linewidth=0.4}\psline(5,6)(8.25,8)}
  \rput(9,2.667){$10$}
  \psframe(8.25,2.167)(9.6,3.167)
  {\psset{linewidth=0.4}\psline(5,2)(8.25,2.667)}
  \rput(9,5.333){$01$}
  \psframe(8.25,4.667)(9.6,5.833)
  {\psset{linewidth=0.4}
  \psline(5,6)(8.25,5.333)}
  \rput(0,0){}
  {\psset{linewidth=0.2,linecolor=gray}\pscurve{<->}(9.7,7.6)(10.2,6.6667)(9.7,5.7333)}
  \rput(9,0){$11$}
  \psframe(8.25,-0.5)(9.6,0.5)
  \psline(5,2)(8.25,0)
  \rput(0,0){}
  {\psset{linewidth=0.2,linecolor=gray}\pscurve{<->}(9.7,2.2667)(10.2,1.3333)(9.7,0.4)}
  \rput(13.5,10){$000$}
  \psframe(12.7,9.5)(14.3,10.5)
  \psline(9.6,8)(12.7,10)
  \rput(13.5,-2){$111$}
  \psframe(12.7,-2.5)(14.3,-1.5)
  \psline(9.6,0)(12.7,-2)
  \rput(13.5,-0.2857){$110$}
  \psframe(12.7,-0.7857)(14.3,0.2143)
  \psline(9.6,0)(12.7,-0.2857)
  \rput(13.5,1.4286){$101$}
  \psframe(12.7,0.9286)(14.3,1.9286)
  {\psset{linewidth=0.4}\psline(9.6,2.667)(12.7,1.4286)}
  \rput(13.5,3.1429){$100$}
  \psframe(12.7,2.6429)(14.3,3.6429)
  {\psset{linewidth=0.4}\psline(9.6,2.667)(12.7,3.1429)}
  \rput(13.5,4.857){$011$}
  \psframe(12.7,4.357)(14.3,5.357)
  \psline(9.6,5.333)(12.7,4.857)
  \rput(13.5,6.571){$010$}
  \psframe(12.7,6.071)(14.3,7.071)
  {\psset{linewidth=0.4}\psline(9.6,5.333)(12.7,6.571)}
  \rput(13.5,8.2857){$001$}
  {\psset{linewidth=0.4}\psline(9.6,8)(12.7,8.2857)}
  \psframe(12.7,7.7857)(14.3,8.7857)
  {\psset{linewidth=0.2,linecolor=gray}\pscurve{<->}(14.5,10)(15.4,9.14285)(14.5,8.2857)}
  \rput(0,0){}
  {\psset{linewidth=0.2,linecolor=gray}\pscurve{<->}(14.5,6.5713)(15.4,5.71415)(14.5,4.857)}
  \rput(0,0){}
  {\psset{linewidth=0.2,linecolor=gray}\pscurve{<->}(14.5,3.1426)(15.4,2.28545)(14.5,1.4283)}
  \rput(0,0){}
  {\psset{linewidth=0.2,linecolor=gray}\pscurve{<->}(14.5,-0.2861)(15.4,-1.14325)(14.5,-2)}
  \rput(0,0){}
  \rput(0.5,5.2){\LARGE C}
  \rput(4.5,3.2){\LARGE D}
  \rput(9,3.9){\LARGE E}
\end{pspicture}
\end{center}
\caption{The input transformation $C\mapsto\mathcal{I}_\sigma(C)$ for $\sigma(0)=1$, $\sigma(1)=0$.
}
\label{FigIPermutation}
\end{figure}

The action of an input transformation is depicted in Figure~\ref{FigIPermutation}:
Changing e.g. the last bit of the input causes a permutation of the outputs corresponding to neighboring branches.
As long as $\Phi$ is complete and closed with respect to that input transformation, the
emulation structure will not be changed. This is a byproduct of the proof of the following theorem:

\begin{theorem}[Input Symmetry]
\label{TheInputSymmetry}
Let $\Phi\subset\Xi$ be positive recurrent, complete and closed with respect to an
input transformation $\mathcal{I}_\sigma$. Then, for every $k\geq |\sigma|$
\[
   \mu([C]_k|\Phi)=\mu([\mathcal{I}_\sigma(C)]_k|\Phi).
\]
\end{theorem}
{\bf Proof.} Suppose that $[C]_k\emu{0}[C_0]_k$, i.e. $C(0\otimes x)=C_0(x)$ for every
$|x|\geq k$, $C\in [C]_k$ and $C_0\in[C_0]_k$. As $|\sigma|\leq k$,
\begin{eqnarray*}
\left(\mathcal{I}_\sigma(C)\right)(0\otimes x)&=&C(\mathcal{I}_\sigma(0\otimes x))=C(0\otimes \mathcal{I}_\sigma
(x))\\
&=&C_0(\mathcal{I}_\sigma(x))=\mathcal{I}_\sigma(C_0)(x),
\end{eqnarray*}
so $[\mathcal{I}_\sigma(C)]_k\emu{0}[\mathcal{I}_\sigma(C_0)]_k$. Analogously,
from $[C]_k\emu{1}[C_1]_k$ it follows that $[\mathcal{I}_\sigma(C)]_k\emu{1}[\mathcal{I}_\sigma(C_1)]_k$
and vice versa. Thus,
\[
   \left(\mathcal{E}_{\Phi,k}\right)_{i,j}=\mu_{[C_i]_k}^{(1)}([C_j]_k|\Phi)
   =\mu_{[\mathcal{I}_\sigma(C_i)]_k}^{(1)}([\mathcal{I}_\sigma(C_j)]_k|\Phi).
\]
So interchanging every equivalence class of computers with its transformed
class leaves the emulation matrix invariant. A similar argument as in
Theorem~\ref{TheOutputSym} proves the claim.\qed

We are now heading towards an analogue of Equation~(\ref{EqProbIdea}), i.e. towards a proof
that our algorithmic string probability equals the weighted average of output frequency. This needs some preparation:
\begin{definition}[Input Symmetry Group]
Let $\mathcal{I}_\sigma$ be an input transformation of order $n\in\N$. A computer $C\in\Xi$
is called {\em $\mathcal{I}_\sigma$-symmetric} if $\mathcal{I}_\sigma(C)=C$ (which is
equivalent to $[\mathcal{I}_\sigma(C)]_n=[C]_n$). The {\em input symmetry group} of $C$
is defined as
\[
   \mathcal{I}-{\rm SYM}(C):=\{\mathcal{I}_\sigma\mbox{ input transformation}\,\,|\,\,
   \mathcal{I}_\sigma(C)=C\}.
\]
\end{definition}
Every transformation of order $n\in\N$ can also be interpreted as a transformation
on $\{0,1\}^N$ for $N>n$, by setting
\[
   \sigma(x_1\otimes x_2\otimes\ldots\otimes x_N):=(x_1\otimes \ldots x_{N-n})
   \otimes \sigma(x_{N-n+1},\ldots,x_N)
\]
whenever $x_i\in\{0,1\}$. With this identification, $\mathcal{I}-{\rm SYM}(C)$ is a group.

\begin{proposition}[Input Symmetry and Irreducibility]
Let $\Phi\subset\Xi$ be irreducible. Then $\mathcal{I}-{\rm SYM}(C)$ is the same
for every $C\in\Phi$ and can be denoted $\mathcal{I}-{\rm SYM}(\Phi)$.
\end{proposition}
{\bf Proof.} Let $\Phi\subset\Xi$ be irreducible, and let $C\in\Phi$ be
$\mathcal{I}_\sigma$-symmetric, i.e. $C(\mathcal{I}_\sigma(s))=C(s)$ for every $s\in\s$.
Let $D\in\Phi$ be an arbitrary computer. Since $\Phi$ is irreducible, it holds
$C\emu{}D$, i.e. there is a string $x\in\s$ with $C(x\otimes s)=D(s)$ for every $s\in\s$.
Let $|s|\geq|\sigma|$, then
\[
   D(s)=C(x\otimes s)=C(\mathcal{I}_\sigma(x\otimes s))=C(x\otimes\mathcal{I}_\sigma(s))
   =D(\mathcal{I}_\sigma(s))
\]
and $D$ is also $\mathcal{I}_\sigma$-symmetric.\qed

For most irreducible computer sets like $\Phi=\Xi^U$, the input symmetry group will
only consist of the identity, i.e. $\mathcal{I}-{\rm SYM}(\Phi)=\{{\rm Id}\}$.

Now we are ready to state the most interesting result of this section:
\begin{theorem}[Equivalence of Definitions]
\label{TheEqDef}
If $\Phi\subset\Xi$ is positive recurrent, complete and closed with respect
to every input transformation $\mathcal{I}_\sigma$ with $|\sigma|\leq n\in\N_0$, then
\[
   \mu(s|\Phi)=\sum_{U\in\Phi}\mu(U|\Phi)\mu_U^{(n)}(s)\mbox{ for every }s\in\bars,
\]
where $\mu_U^{(n)}(s)$ is the output frequency as introduced in Definition~\ref{DefAlgFrequency}.
\end{theorem}
{\bf Proof.} The case $n=0$ is trivial, so let $n\geq 1$. It is convenient to introduce another
equivalence relation on the computer classes. We define the corresponding equivalence classes
(``transformation classes'') as
\[
   \{V\}_k:=\left\{[X]_k\in[\Phi]_k\,\,|\,\,\exists \mathcal{I}_\sigma:|\sigma|\leq k,
   [\mathcal{I}_\sigma(V)]_k=[X]_k\right\}.
\]
Thus, two computer classes $[X]_k$ and $[Y]_k$ are elements of the same transformation class
if one is an input transformation (of order less than $k$) of the other. Again, we set
$\{\Phi\}_k:=\{\{X\}_k\,\,|\,\,X\in\Phi\}$.

For every $X\in[X]_n$, the probability $\mu_X^{(n)}(s|\Phi)$ is the same and can be denoted
$\mu_{[X]_n}^{(n)}(s|\Phi)$. According to Proposition~\ref{EqCKString}, we have
\begin{eqnarray*}
   \mu(s|\Phi)=\sum_{\{X\}_n\in\{\Phi\}_n} \sum_{[Y]_n\in\{X\}_n}
   \mu([Y]_n|\Phi)\mu_{[Y]_n}^{(n)}(s|\Phi).
\end{eqnarray*}
Due to Theorem~\ref{TheInputSymmetry}, the probability $\mu([Y]_n|\Phi)$ is the same
for every $[Y]_n\in\{X\}_n$. Let $[X]_n$ be an arbitrary representative of $\{X\}_n$, then
\[
  \mu(\{X\}_n|\Phi):=\sum_{[Y]_n\in\{X\}_n} \mu([Y]_n|\Phi)=\# \{X\}_n \cdot
  \mu([X]_n|\Phi).
\]
The two equations yield
\[
   \mu(s|\Phi)=\sum_{\{X\}_n\in\{\Phi\}_n} \frac{\mu(\{X\}_n|\Phi)}{\#\{X\}_n}
   \sum_{[Y]_n\in\{X\}_n} \mu_{[Y]_n}^{(n)}(s|\Phi).
\]
Let $\mathbf{S}_{2^n}$ be the set of all permutations on $\{0,1\}^n$. Two permutations
$\sigma_1,\sigma_2\in\mathbf{S}_{2^n}$ are called $\Phi$-equivalent if there exists
a $\sigma\in\mathcal{I}-{\rm SYM}(\Phi)$ such that $\sigma_1=\sigma\circ\sigma_2$
(recall that $\Phi$ is irreducible). This is the case if and only of $\mathcal{I}_{\sigma_1}(C)
=\mathcal{I}_{\sigma_2}(C)$ for one and thus every computer $C\in\Phi$. The set of all $\Phi$-equivalence classes
will be denoted $\mathbf{S}_n(\Phi)$. Every computer class $[Y]_n\in\{X\}_n$ is generated from $[X]_n$ by
some input transformation. If $X$ is an arbitrary representative of $[X]_n$, we thus have
\[
   \mu(s|\Phi)=\sum_{\{X\}_n\in\{\Phi\}_n} \frac{\mu(\{X\}_n|\Phi)}{\#\{X\}_n}
   \sum_{[\sigma]\in\mathbf{S}_n(\Phi)}\mu_{[\mathcal{I}_\sigma(X)]_n}^{(n)}(s|\Phi),
\]
where $\sigma\in[\sigma]$ is an arbitrary representative.
For every equivalence class $[\sigma]$, it holds true
$\#[\sigma]=\#(\mathbf{S}_{2^n} \cap \mathcal{I}-{\rm SYM}(\Phi))$, thus
\begin{eqnarray*}
   \mu(s|\Phi)=\sum_{\{X\}_n\in\{\Phi\}_n}& \frac{\mu(\{X\}_n|\Phi)}{\#\{X\}_n}
   \cdot \frac 1 {\#(\mathcal{I}{\rm-SYM}(\Phi)\cap \mathbf{S}_{2^n})}\\
   &\sum_{\sigma\in\mathbf{S}_{2^n}}\mu_{[\mathcal{I}_\sigma(X)]_n}^{(n)}(s|\Phi).
\end{eqnarray*}
By definition of the set $\mathbf{S}_n(\Phi)$,
\[
   \# \{X\}_n\cdot\#(\mathcal{I}{\rm-SYM}(\Phi)\cap \mathbf{S}_{2^n})=
   \# \mathbf{S}_{2^n}=(2^n)!.
\]
Using that $\#\{X\}_n=\#\mathbf{S}_n(\Phi)$, we obtain
\begin{eqnarray*}
   \mu(s|\Phi)&=&\sum_{\{X\}_n\in\{\Phi\}_n}\frac{\mu(\{X\}_n|\Phi)}{(2^n)!}
   \sum_{\sigma\in\mathbf{S}_{2^n}}\mu_{\mathcal{I}_\sigma(X)}^{(n)}(s|\Phi)\\
   &=&\sum_{\{X\}_n\in\{\Phi\}_n}\frac{\mu(\{X\}_n|\Phi)}{(2^n)!}
   \sum_{\sigma\in\mathbf{S}_{2^n}}\\
   &&\qquad\qquad\sum_{x\in\{0,1\}^n} \delta_{\mathcal{I}_\sigma(X)(x),s}
   \,\mu_{X^{-1}(\Phi)}(x).
\end{eqnarray*}
As $|x|=n\geq|\sigma|$ it holds $\mathcal{I}_\sigma(X)(x)=X(\mathcal{I}_\sigma(x))=X(\sigma(x))$.
The substitution $y:=\sigma(x)$ yields
\begin{eqnarray*}
   \mu(s|\Phi)&=&\sum_{\{X\}_n\in\{\Phi\}_n}\frac{\mu(\{X\}_n|\Phi)}{(2^n)!}
   \sum_{y\in\{0,1\}^n} \delta_{X(y),s}\\
   &&\qquad\qquad\qquad\sum_{\sigma\in\mathbf{S}_{2^n}} \mu_{X^{-1}(\Phi)}
   (\sigma^{-1}(y)).
\end{eqnarray*}
Up to normalization, the rightmost sum is the average of all permutations of
the probability vector $\mu_{X^{-1}(\Phi)}$, thus
\[
   \frac 1 {(2^n)!} \sum_{\sigma\in\mathbf{S}_{2^n}} \mu_{X^{-1}(\Phi)}(\sigma^{-1}(y))=2^{-n}.
\]
Recall that $X$ was an arbitrary representative of an arbitrary representative of $\{X\}_n$.
The last two equations yield
\begin{eqnarray*}
   \mu(s|\Phi)&=&\sum_{\{X\}_n\in\{\Phi\}_n}
   \mu(\{X\}_n|\Phi)\sum_{y\in\{0,1\}^n} \delta_{X(y),s}2^{-n}\\
   &=&\sum_{\{X\}_n\in\{\Phi\}_n}\sum_{[X]_n\in\{X\}_n}
   \mu([X]_n|\Phi)\mu_x^{(n)}(s)\\
   &=&\sum_{\{X\}_n\in\{\Phi\}_n}\sum_{[X]_n\in\{X\}_n}\sum_{X\in[X]_n}
   \mu(X|\Phi)\mu_x^{(n)}(s)\\
   &=&\sum_{X\in\Phi}\mu(X|\Phi)\mu_X^{(n)}(s).
\end{eqnarray*}
Note that if $X$ and $Y$ are representatives of representatives of an arbitrary transformation class
$\{X\}_n$, then $\mu_X^{(n)}(s)=\mu_Y^{(n)}(s)$.\qed

This theorem is the promised analogue of Equation~(\ref{EqProbIdea}): it shows that
the string probability that we have defined in Definition~\ref{DefStringProbability} is the weighted average of output frequency
as defined in Definition~\ref{DefAlgFrequency}.
For a discussion why this is interesting and surprising, see the first few paragraphs of this appendix.

\section*{Acknowledgments}
The author would like to thank
N. Ay, D. Gross, S. Guttenberg, M. Ioffe, T. Kr\"uger, D. Schleicher, F.-J. Schmitt, R. Siegmund-Schultze, R. Seiler,
and A. Szko\l a for helpful discussions and kind support.

\end{document}